\documentclass[aps,prd,showpacs,floatfix,notitlepage,nofootinbib,twocolumn,a4paper]{revtex4}
\usepackage{amsfonts,amsmath,units,wasysym,epsfig,graphicx,verbatim,color,subfigure,graphicx,bm,mathrsfs,lipsum,hyperref,cleveref}
\usepackage{booktabs}
\usepackage[normalem]{ulem}  
\definecolor{dgreen}{RGB}{0, 204, 0}
\begin{document}
\newcommand{\bhaskar}[1]{\textcolor{red}{ \bf BB: #1}}
\newcommand{\sukanta}[1]{{\textcolor{magenta}{\bf{SB: #1}} }}
\newcommand{\pc}[1]{\textsf{\color{blue}{ PC: #1}}}
\newcommand{\rn}[1]{\textcolor{dgreen}{ \bf RN: #1}}

\newcommand{\IUCAA}{Inter-University Centre for Astronomy and Astrophysics, Post Bag 4, Ganeshkhind, Pune 411 007, India}
\newcommand{\INFN}{INFN Sezione di Ferrara, Via Saragat 1, 44122 Ferrara, Italy} 
\newcommand{\LIEGE}{Space  sciences,  Technologies  and  Astrophysics  Research  (STAR)  Institute,  Université de Liège, Bât. B5a, 4000 Liège, Belgium} 
\newcommand{\PM}{Polba Mahavidyalaya, Hooghly, West Bengal 712148, India}  
\newcommand{\WSU}{Department of Physics \& Astronomy, Washington State University, 1245 Webster, Pullman, WA 99164-2814, U.S.A}

\title{Towards mitigation of apparent tension between nuclear physics and astrophysical observations by improved modeling of neutron star matter}


\author{Bhaskar Biswas $^{\rm 1}$, Prasanta Char $^{\rm 2, 3}$, Rana Nandi $^{\rm 4}$,  Sukanta Bose $^{\rm 1, 5}$}
\affiliation{$^{\rm 1}$ \IUCAA,$^{\rm 2}$ \INFN,$^{\rm 3}$ \LIEGE, $^{\rm 4}$ \PM,$^{\rm 5}$\WSU}

\begin{abstract}
Observations of neutron stars (NSs) by the LIGO-Virgo and NICER collaborations have provided reasonably precise measurements of their various macroscopic properties. In this paper, we employ a Bayesian framework to combine them and place improved joint constraints on the properties of NS
equation of state (EoS). We use a hybrid EoS formulation that  employs a parabolic expansion-based nuclear empirical parameterization around the nuclear saturation density augmented by a generic 3-segment piecewise polytrope model at higher densities. Within the $90 \%$ credible level this parameterization predicts $R_{1.4} = 12.57_{-0.92}^{+0.73}$ km and $\Lambda_{1.4} = 550_{-225}^{+223}$ for the radius and dimensionless tidal deformability, respectively, of a $1.4 M_{\odot}$ NS. Finally, we show how the construction of the full NS EoS based solely on the nuclear empirical parameters at saturation density leads to certain tension with the astrophysical data, and how the hybrid approach provides a resolution to it.

\end{abstract}

\preprint{LIGO-P2000221}

\maketitle

\section{Introduction}The properties and composition of nuclear matter near and above nuclear saturation density ($\rho_0$) have been the subject of intense theoretical and experimental investigations throughout the preceding decades. Fascinatingly, important insights about them can be deduced from observations of macroscopic properties of NSs, such as their mass, radius, moment of inertia, and tidal deformability. After all, it is likely that the NS interior hosts matter at densities reaching supranuclear values. Recent simultaneous mass and radius measurements of PSR J0030+0451 by NICER collaboration \cite{Riley:2019yda,Miller:2019cac}, the mass measurements of the pulsars PSR J0348+0432 \cite{Antoniadis:2013pzd}, PSR J0740+6620 \cite{Cromartie:2019kug} exceeding $2M_\odot$, and the binary neutron star (BNS) merger events GW170817~\cite{TheLIGOScientific:2017qsa} and GW190425~\cite{Abbott:2020uma} reported by the LIGO/VIRGO collaboration~\cite{advanced-ligo,advanced-virgo} (LVC)  have provided an extraordinary amount of information about NS composition. 
These individual measurements can be combined to impose very strict constraints on the EoS of such  matter \cite{Abbott:2018exr,Raaijmakers:2019dks,Jiang:2019rcw,Landry:2020vaw}. 

In the literature, one can find several strategies to approximate nuclear EoSs, such as spectral parameterizations and the piecewise-polytrope (PP) approximation~\cite{Jiang:2019rcw,Read:2008iy,Lindblom:2010bb}. Both approximations were constructed to match a wide variety of theoretical EoSs with only a few parameters. However, due to the small number of parameters employed, the fitting procedure introduces some errors~\cite{Lackey-2015,Carney:2018sdv}. Although the spectral representation does better in terms of reducing the fitting error in comparison with the PP model, a faithful reproduction of the full range of EoS variability is not possible, in general, by either of them. A different approach worth discussing is the non-parametric inference of the EoS \cite{Landry:2018prl,Essick:2019ldf, Landry:2020vaw}. In this method, a large number of EoS functionals are generated whose ranges in the pressure-density plane are loosely guided by a certain number of widely used candidate nuclear EoSs from the literature, without the explicit need for any type of parameterization. 
 Furthermore, this generation process can be adapted to reproduce features of the candidate EoSs, therein incorporating information from nuclear physics. Other approaches motivated from nuclear physics including chiral effective field theory calculations~\cite{Capano:2019eae,Raaijmakers:2019dks,Essick-2020arXiv} at low densities combined with agnostic parameterization at higher densities have also been explored in several works. There has been some uncertainties on the range up to which density the chiral EFT calculation can reproduce nuclear properties. Recently, it is suggested~\cite{Essick-2020arXiv} from astrophysical data that up to $\sim 2\rho_0$, the chiral EFT is favored over other generic approaches.

In this paper instead of using chiral EFT prediction we propose an alternative method: at low densities we use the parabolic expansion of the binding energy per nucleon of nuclear matter about $\rho_0$. The coefficients of the expansion, known as the nuclear empirical parameters, can be constrained systematically by combining prior knowledge from nuclear physics and astrophysical observations \cite{Piekarewicz:2008nh}. Then at high densities where EoS is completely unknown from our nuclear physics knowledge, we use a three-piece PP model. In Ref. \cite{Steiner:2010fz}, a similar type of model was used to infer NS properties from low mass X-ray binary (LMXB) data. We differ from their approach by using a three-segment polytrope instead of a two-segment one, and also a fixed transition density between the nuclear physics informed and agnostic model that we select from Bayesian evidence calculation.

At this point, we would like to emphasize a critical aspect of the use of nuclear empirical parameterization in the literature. One can find several works~\cite{Guven:2020dok,Zhang:2019fog,Xie:2019sqb,dEtivaux:2019cnf,Carson2019:nucl170817,Zimmerman:2020eho} featuring this model and its characteristics such as correlation between expansion parameters, and extensive studies of NS properties. But, the validity of this model at all densities in the interior of a NS is questionable. Since the expansion is about $\rho_0$, there can be large modelling errors if this is applied  to densities as high as $\sim 5-6 \rho_0$ at the core of a NS. Therefore, one should be careful while implementing this model to neutron stars for the purpose of inferring nuclear matter properties. Recently, some empirical parameters like the slope of symmetry energy and its derivative have been studied in the light of combined constraints \cite{Zimmerman:2020eho,Xie:2020tdo}. In these studies, the complete posterior distribution of the measurements were not used. Particularly, Ref.~\cite{Xie:2020tdo} has used several marginalized radius distributions which, in principle, can add some biases to inferred values of EoS parameters. On the other hand, Ref.~\cite{Zimmerman:2020eho} utilized correlations amongst certain combinations of nuclear parameters and the radius of a $1.4  M_{\odot}$ NS, or the tidal deformability, to deduce constraints for a limited set of nuclear EoSs. In Ref.~\cite{Guven:2020dok}, an apparent tension has also been found between the empirical parameter based modelling of neutron stars and astrophysical observations. We show in this article that such discrepancies arise due to certain deficiencies in EoS modeling, especially, at high densities. We also improve upon certain other aspects of past studies as well: instead of using theoretically derived correlations, which can have an inherent bias due to the choice of the EoSs, we sample the posterior space of the EoS parameters directly using the data with the Nested Sampling algorithm \cite{Skilling_Nested}. 

The paper is organized as follows. In section \ref{section:EoS}, we present details of the construction of the EoS. Then, we briefly describe our Bayesian formalism in section \ref{section:Bayes}. We present the results obtained combining the relevant publicly available astrophysical measurements in section \ref{section:results}. Here, we particularly discuss the macroscopic properties of neutron stars in \ref{macro-prop}, followed by a comparison with a situation where only PP parameterization is used to model the star in \ref{comparePP}. After that we address the issue of ``tension" arising due to improper modelling in \ref{tension}. Finally, we summarize and conclude in section \ref{summary}.

\section{Equation of state}
\label{section:EoS}
The equation of state of nuclear matter around $\rho_0$ is described in terms of the nuclear empirical parameters \cite{Piekarewicz:2008nh,Margueron:2017eqc}. These are defined from the parabolic expansion of the energy per nucleon  $e(\rho,\delta)$ of asymmetric nuclear matter as:
\begin{equation}
    e(\rho,\delta) \approx  e_0(\rho) +  e_{\rm sym}\delta^2,
\end{equation}
where $e_0(\rho)$ is the energy per nucleon in symmetric nuclear matter
that contains equal numbers of neutrons and protons,
$e_{\rm sym}(\rho)$ is the nuclear symmetry energy, the energy cost for
having asymmetry in the number of neutrons and protons in the system
and $\delta=(\rho_n-\rho_p)/\rho$ is the measure of this asymmetry. 
We parameterize $e_0(\rho)$ and $e_{\rm sym}(\rho)$ around the
saturation density $\rho_0$ as:
\begin{eqnarray}
 e_0(\rho) &=&  e_0(\rho_0) + \frac{ K_0}{2}\chi^2 + \frac{ J_0}{6}\chi^3\label{eq:e0} +\,...,\\
e_{\rm sym}(\rho) &=&  e_{\rm sym}(\rho_0) + L\chi + \frac{ K_{\rm sym}}{2}\chi^2 
+ \frac{J_{\rm sym}}{6}\chi^3 + ..., \label{eq:esym}
\end{eqnarray}
where $\chi \equiv (\rho-\rho_0)/3\rho_0$ quantifies deviation from saturation density.
In this work, we assume the NS interior to be composed solely of nucleonic matter whose properties can be extrapolated from the saturation characteristics embodied in the nuclear empirical parameters introduced above. A review of the experimental determination and theoretical estimation of these parameters can be found in Ref.~\cite{Margueron:2017eqc}.

\begin{table}[ht]
    \centering
    \begin{tabular}{c|c}
         \hline
         Parameter& Prior  \\
         \hline
         $  K_0$ (MeV) & $\mathcal{N}(240,30)$ \\
         $ e_{\rm sym}$ (MeV) & $\mathcal{N}(31.7,3.2)$ \\
         $L$ (MeV) & $\mathcal{N}(58.7,28.1)$ \\
         $ K_{\rm sym}$ (MeV) & uniform(-400,100) \\
         $ \Gamma_1$  & uniform(1.4,5)\\
         $ \Gamma_2$  & uniform(1,5)\\
         $ \Gamma_3$  & uniform(1.,5)\\
         \hline
    \end{tabular}
    \caption{Prior ranges of various EoS parameters.}
    \label{tab:prior}
\end{table}

The uncertainties in the lower order parameters are quite small, and are fairly well determined~\cite{Brown:2013pwa,Tsang:2019ymt}. 
Hence, we keep the lowest-order empirical parameters fixed, such as $e_0 (\rho_0) = -15.9$ MeV and $\rho_0= 0.16$ fm$^{-3}$. 
At the next order, the parameters incorporated are the curvature of symmetric matter ($K_0$), the symmetry energy ($e_{\rm sym}$) and slope ($L$) of that energy at $\rho_0$.
While their uncertainties are larger, plenty of experimental data exist that constrain their ranges~\cite{Piekarewicz:2009gb,Lattimer:2012xj,Oertel:2016bki,  Zhang:2019fog}.
Therefore, for the Bayesian inference, their priors are taken to be Gaussian distributions with 
spreads set to those ranges and means set to their medians.
The higher-order parameters are not well-constrained but are necessary for understanding the high-density behavior inside the NS. Therefore, we choose large ranges of uniformly probable values for the curvature of symmetry energy ($K_{\rm sym}$) and the skewness ($J_0$, $J_{\rm sym}$) as their priors~\cite{Kolomeitsev:2016sjl,Margueron:2017eqc,Margueron:2018eob}. 

Due to the lack of theoretical understanding of nuclear matter at the supra-nuclear density we use a three-piecewise polytrope parameterization ($\Gamma_1, \Gamma_2, \Gamma_3$) with fixed transition densities. We create a set of models (see the Bayesian methodology described in section~\ref{section:Bayes}) with a varying transition density between $1.1-1.7 n_0$ where the first polytrope is attached. Then we compare each model with the data following the interpretation of ~\citet{Bayes-factor}, but do not find  significant evidence for any particular transition density. However, we observe the evidence is slightly higher around $1.25n_0$ and therefore, fix the first transition point there. Similarly, Ref~\cite{Essick-2020arXiv} found a local maximum in the evidence for theoretical models below $2\rho_0$. We use the first polytrope between $1.25$ to $1.8 n_0$. Then we stitch the second polytrope upto $3.6 n_0$. We use the last polytrope above that density.

\begin{figure}[ht]
    \centering
    \includegraphics[width=.45\textwidth]{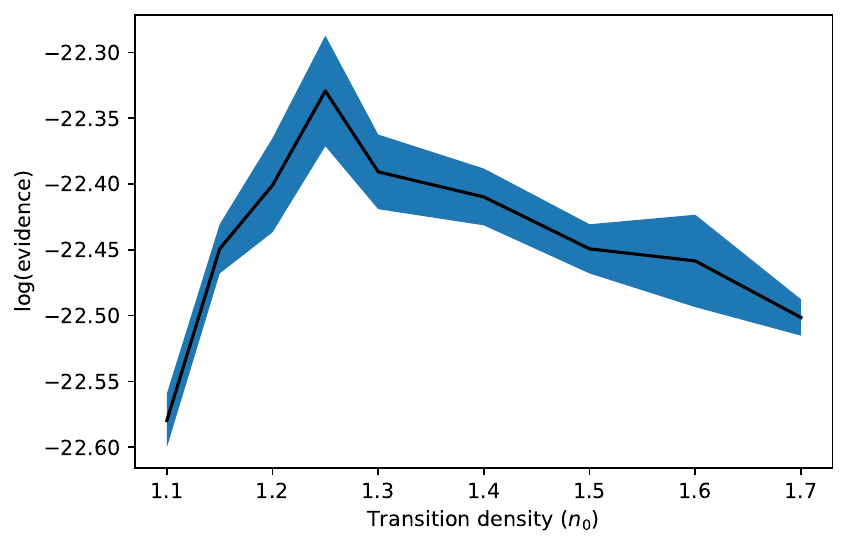}
    \caption{Bayesian Evidence of the astrophysical data is compared between a set of nested models with varying transition densities where the first polytrope is attached. We find 1.25 $n_0$ is the most favored transition density. }
    \label{fig:evidence}
\end{figure}

In our hybrid nuclear+PP model, we truncate the Taylor expansion in $e_0 (\rho_0)$ and $e_{\rm sym} (\rho_0)$ upto second order in $\chi$. These prior ranges are listed in Table~\ref{tab:prior}. Additionally, we
require the EoS to be causal throughout the density range, which implies
that the speed of sound never exceeds the speed of light. Finally, as the choice of the crustal EoS does not significantly influence the NS observables \cite{Biswas:2019ifs,Gamba:2019kwu}, we use the standard BPS crust \cite{1971ApJ...170..299B} for the low density regime  and join it with the high density EoS in a consistent fashion as described in Ref. \cite{Xie:2019sqb} . 
 
\section{Bayesian methodology}
\label{section:Bayes}
We discuss the Bayesian methodology that is used in this work to construct the posteriors of the EoS parameters of NSs by combining astrophysical data from PSR J0740+6620, GW170817, GW190425, and NICER observations. Our Bayesian methodology is similar to some previous works~\cite{Lackey-2015,Landry:2020vaw,Raaijmakers:2019dks}. In GW observations, the chirp mass $\mathcal{M}_c$ and the effective tidal deformability $\widetilde{\Lambda}$ (which is a mass-weighted combination of the deformabilities of the two components of a binary)
carry information about the EoS. For the NICER observations, the mass-radius posterior of the sources obtained by pulse-profile modelling directly provide  information about the EoS.   
Therefore, the posterior of the EoS parameters can be written as,
\begin{equation}
    P(\theta | {d}) = \frac{P ({d} | \theta) \times P(\theta)}{P(d)}\, = \frac{\Pi_i P ({d_i} | \theta) \times P(\theta)}{P(d)}\,,
    \label{bayes theorem}
\end{equation}
where $\theta = ( K_0 , e_{\rm sym}, L, K_{\rm sym}, \Gamma_1, \Gamma_2, \Gamma_3)$ is the set of our EoS parameters, $d = (d_{\rm GW}, d_{\rm X-ray}, d_{\rm Radio})$ is the set of data from the three different 
types of observations that are
used to construct the likelihood, $P(\theta)$ are the priors of those parameters and $P(d)$ is the Bayesian evidence, given the particular EoS model. For GW observations, information about EoS parameters come from the masses $m_1, m_2$ of the two binary components and the corresponding tidal deformabilities $\Lambda_1, \Lambda_2$. In this case,
\begin{align}
    P(d_{GW}|\theta) = \int^{M_{\rm max }(\theta)}dm_1 \int^{m_1} dm_2 \int d\Lambda_1 \int d\Lambda_2        \nonumber \\ \delta(\Lambda_1 - \Lambda_1(\theta, m_1))
    \delta(\Lambda_1 - \Lambda_2(\theta, m_2)) 
     \nonumber \\
    P(d_{GW} | m_1, m_2, \Lambda_1, \Lambda_2) \,,
\end{align}
where $M_{\rm max }(\theta)$ is the maximum mass of a NS for a particular set of EoS parameter $\theta$. Given the high-precision measurement of the chirp mass in GW observations, we fix it to the observed median value and use it to generate a set of binary neutron star systems within the allowed range of mass-ratios,  which was also determined by observations. Therefore, in our likelihood evaluation, we marginalize over the mass ratio $q=m_2/m_1$ instead of individual masses. We modelled the likelihood with Gaussian kernel density estimator (KDE) based on the publicly available samples. 

X-ray observations give the mass ($m$) and radius ($R$) measurements of NS. Therefore, the corresponding likelihood takes the following form,
\begin{align}
    P(d_{\rm X-ray}|\theta) = \int^{M_{\rm max }(\theta)} dm \int   dR   \delta(R - R(\theta, m))  \nonumber \\
    P(d_{\rm X-ray} | m, R) \,.
\end{align}
Similar to GW observations, we modelled the likelihood for X-ray with KDE.

On the other hand, radio-pulsar observations provide us with very accurate measurements of the NS mass. In our analysis, following~\cite{Miller:2019nzo,Raaijmakers:2019dks,Landry:2020vaw}, we marginalize over the heaviest pulsar mass measurements~\cite{Cromartie:2019kug} taking into account its measurement uncertainties,
\begin{align}
    P(d_{\rm Radio}|\theta) = \int^{M_{\rm max }(\theta)} dm   
    P(d_{\rm Radio} | m) \,.
\end{align}
In order to populate the posterior distribution of Eq.~\ref{bayes theorem}, we implement nested sampling
algorithm by employing the publicly available python based {\tt Pymultinest}~\cite{pymultinest} package.

 In this current work, we assume a flat distribution of NS masses for all the events. Although it may not necessarily be true. When combining information one needs to employ a population model for the observed sources. For binary neutron stars, this population model can be the astrophysical distribution of
NS masses or NS central densities. In order to obtain unbiased
results, this population model needs to be fitted and marginalized simultaneously with the EoS inference. It has been shown~\cite{Wysocki-2020} that this bias will only become important after$\sim 20-30$ observations, so with the current data it is acceptable to simply fix the population
model by employing a flat mass
distribution.

\section{Results from current observations}
\label{section:results}
\begin{figure*}[ht]
    \centering
    \includegraphics[width=\textwidth]{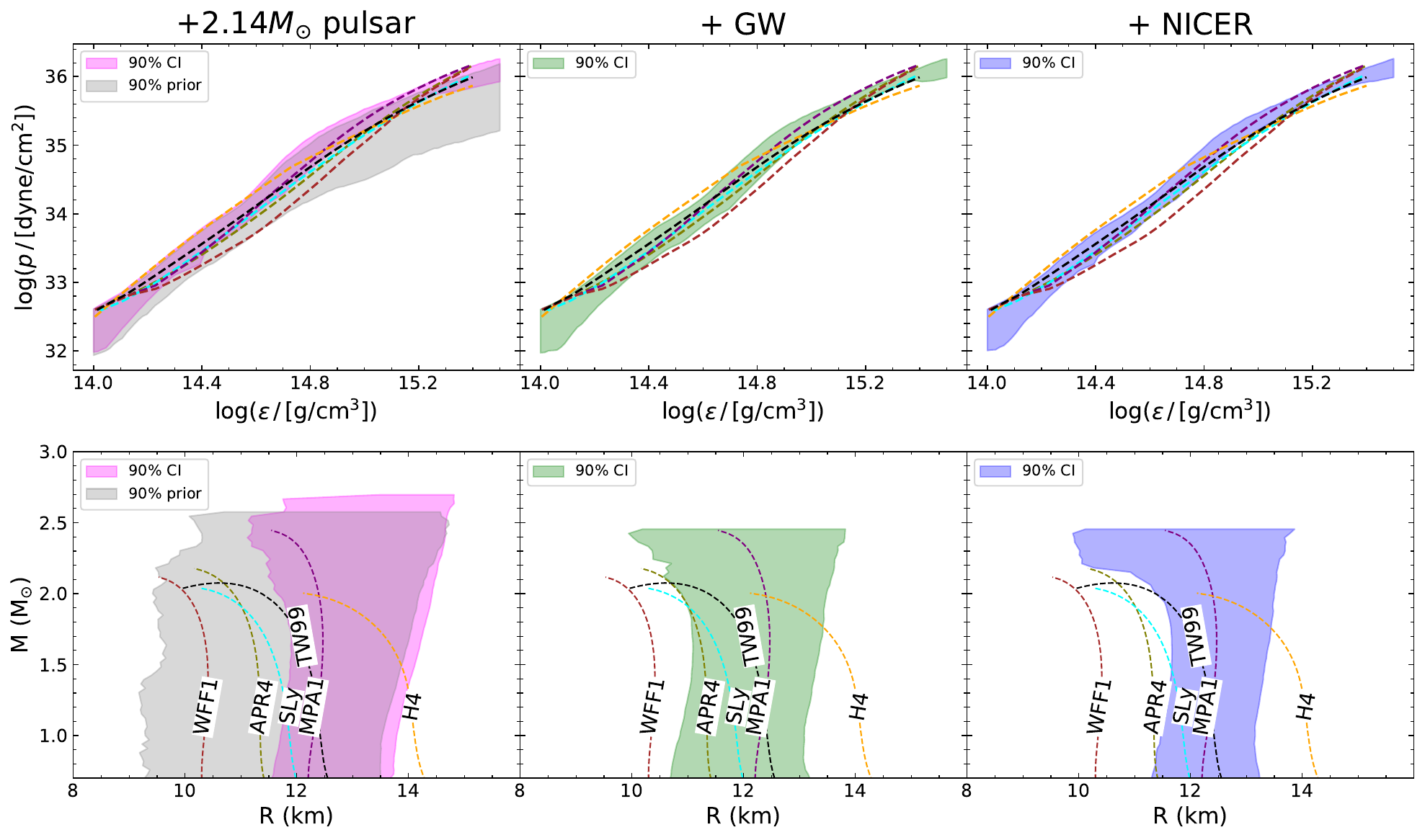}
    \caption{In the top panel, the marginalized posterior distribution of the pressure in NS interior as a function of energy is shown, using nuclear physics informed prior: (i) at left PSR J0740+6620 alone, (ii) in the middle two GW observations are combined and (iii) at right NICER data is added. Some standard EoS curves like APR \cite{1998PhRvC..58.1804A}, SLy \cite{2001A&A...380..151D}, WFF1 \cite{1988PhRvC..38.1010W}, MPA1 \cite{1987PhLB..199..469M}, H4 \cite{2006PhRvD..73b4021L} and TW99 \cite{Typel:1999yq} are also overlaid. 
    In the bottom panel corresponding mass-radius posterior distributions are shown. In left panel (both top and bottom) $90 \%$ CR of prior is shown in grey colour. 
    }
    \label{fig:post-poly}
\end{figure*}
In this paper, we use the Bayesian framework described above to combine neutron star data from radio, GWs, and NICER to construct posteriors of the EoS parameters. We use the mass measurement of PSR J0740+6620 as a Gaussian likelihood with a median of 2.14 $M_{\odot}$ and standard deviation of $0.1 M_{\odot}$. The GW data utilized are the $m_1,m_2,\Lambda_1,\Lambda_2$ distribution of GW170817~\footnote{LVK collaboration,~\href{https://dcc.ligo.org/LIGO-P1800115/public}{https://dcc.ligo.org/LIGO-P1800115/public}} and GW190425\footnote{LVK collaboration,~\href{https://dcc.ligo.org/LIGO-P2000026/public}{https://dcc.ligo.org/LIGO-P2000026/public}}, which are publicly available. For NICER observations we use the 3-spot mass-radius samples~\footnote{Released mass-radius samples released by Miller et al.,~\href{https://zenodo.org/record/3473466\#.XrOt1nWlxBc}{https://zenodo.org/record/3473466\#.XrOt1nWlxBc}} by Miller et al.~\cite{Miller:2019cac}. 

\subsection{Macroscopic properties}
\label{macro-prop}

In the top panel of Fig.~\ref{fig:post-poly}, the resulting marginalized posterior distributions of pressure inside NS are plotted as a function of energy density by adding successive observations, and in the bottom panel corresponding mass-radius posterior distributions are shown. To obtain this plot, 90\% credible region (CR) of pressure (radius) is computed at a fixed energy density (mass) and those are plotted as a function of energy density (mass).
 In the left panel in grey colour we also show the $90 \%$ CR of prior in both EoS and mass-radius distribution. The magenta band is the posterior deduced from PSR J0740+6620 data alone and inspecting the prior distribution we see it mostly favours the stiffer EoSs.

The green band is obtained after adding two GW detections with PSR J0740+6620, favoring relatively softer EoSs than the first case when only constraints from PSR J0740+6620 are used.  Similar constraints were obtained from GW170817 data  by LVC using the spectral EoS representation, as well as by other authors using various other parameterizations~\cite{Landry:2020vaw,Essick-2020arXiv,Coughlin:2018fis,Dietrich:2020lps}.
 In blue, the joint posterior of PSR J0740+6620, NICER, and the two GW observations is plotted; it favors a stiffer EoS than what we obtained in the middle panel.  Addition of NICER reduces the uncertainties in the EoS, and various macroscopic NS properties as a consequence. For example, the uncertainty in the measurement of $R_{1.4}$ is $\sim 2.19$ km by combined PSR J0740+6620 and GW observations alone. This shrinks to  $\sim 1.65$ km when combined with NICER. Ref.~\cite{Essick-2020arXiv} found a similar reduction in the CR of $R_{1.4}$, when they include theoretical information at low densities.

In Table~\ref{tab2}, we report the median and $90 \%$ CR of all EoS parameters, radius, $\Lambda$ of $1.4 M_{\odot}$ NS, and maximum mass $ M_{\rm max}$ by adding successive observations.

\begin{table}[ht]
\begin{tabular}{cccccc} 
\hline
 Quantity& $2.14 M_{\odot}$ pulsar & +GW & +NICER\\

\hline 
\vspace{1ex}
 $K_0$ (MeV) &$242_{-48}^{+48}  $ & $240_{-48}^{+47}$ &  $241_{-47}^{+47}$\\

\vspace{1ex}
 $e_{\rm sym}$ (MeV) &$31.8_{-5.2}^{+5.1}  $ & $32.0_{-5.1}^{+5.1}$ &  $32.0_{-5.0}^{+5.1}$\\

\vspace{1ex}
 $L$ (MeV) &$72.6_{-32.1}^{+33.2}  $ & $58.0_{-28.4}^{+31.8}$ &  $61.2_{-25.2}^{+29.6}$\\

\vspace{1ex}
 $ K_{\rm sym}$ (MeV) &$-106_{-230}^{+176}$ & $-191_{-174}^{+208}$ &  $-181_{-182}^{+204}$\\

 \vspace{1ex}
 $\Gamma_1$  &$3.82_{-1.88}^{+0.92}  $ & $3.21_{-1.57}^{+1.54}$ &  $3.45_{-1.65}^{+1.34}$\\

  \vspace{1ex}
 $\Gamma_2$  &$3.83_{-1.42}^{+0.94}  $ & $3.94_{-0.93}^{+0.72}$ &  $3.92_{-0.94}^{+0.75}$\\

  \vspace{1ex}
 $\Gamma_3$  &$3.13_{-1.87}^{+1.62}  $ & $3.09_{-1.69}^{+1.55}$ &  $3.11_{-1.79}^{+1.59}$\\

\vspace{1ex}

 $R_{1.4} [\rm km]$ &$13.14_{-1.27}^{+0.87}  $ & $12.43_{-1.37}^{+0.82}$ &  $12.57_{-0.92}^{+0.73}$\\

\vspace{1ex}
$\Lambda_{1.4}$ & $739_{-359}^{+357}$ & $509_{-266}^{+267}$ & $550_{-225}^{+223}$ \\

\vspace{1ex}
$M_{\rm max} (M_{\odot})$ & $2.31_{-0.24}^{+0.32}$ & $2.21_{-0.16}^{+0.22}$ & $2.23_{-0.18}^{+0.21}$ \\

\hline 
\end{tabular}
\caption{Median and $90 \%$ CR of nuclear parameters,  $R_{1.4}$, $\Lambda_{1.4}$ and maximum mass $\rm\ M_{max}$ are quoted here. 
}
\label{tab2}
\end{table}

\subsection{Comparison to Piecewise polytrope parameterization}
\label{comparePP}
\begin{figure*}[ht]
    \centering
    \includegraphics[width=\textwidth]{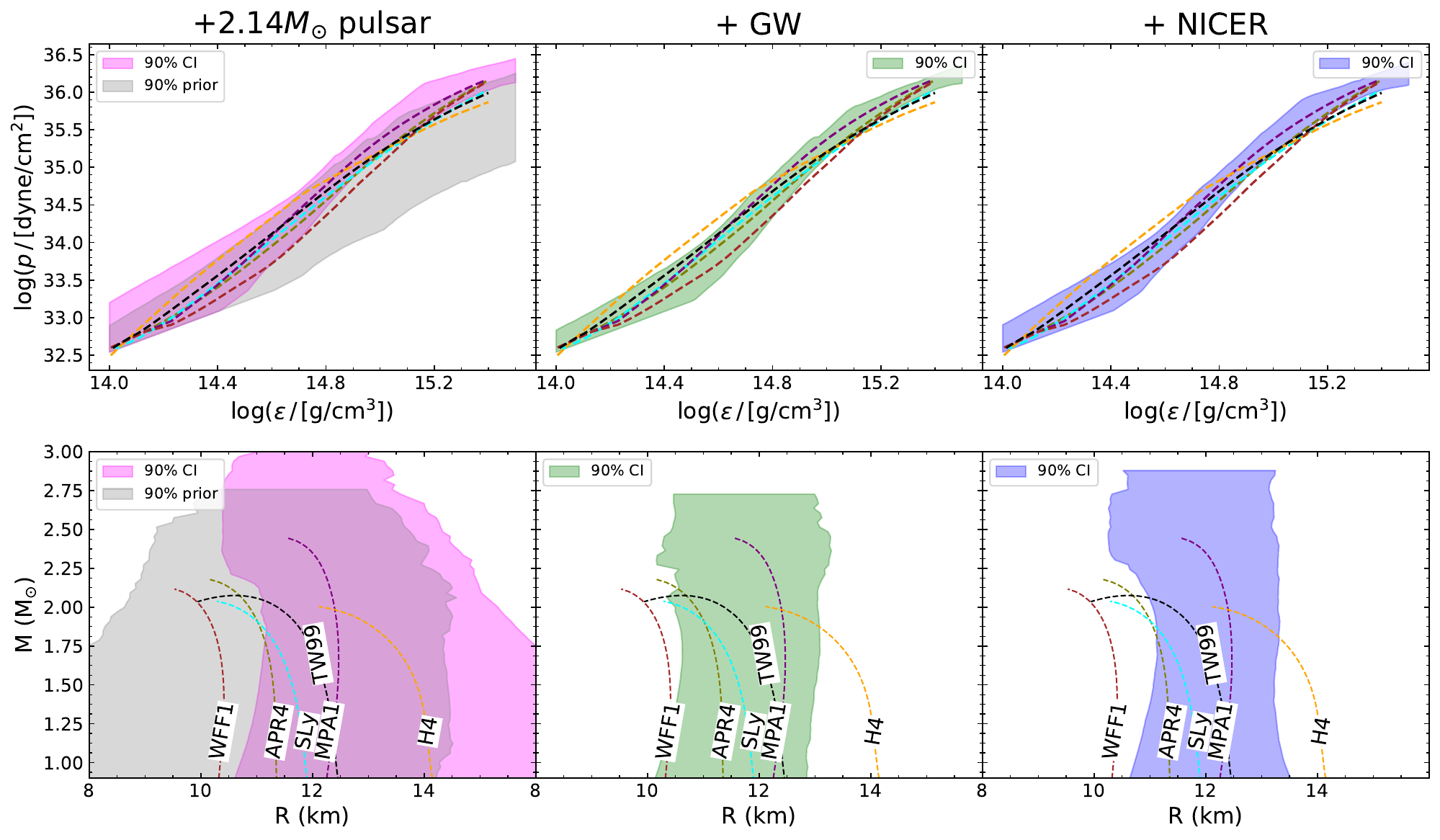}
    \caption{ This figure is similar to Fig.~\ref{fig:post-poly} but with piecewise-polytrope parameterization.
    }
    \label{fig:EoS-post-pp}
\end{figure*}
Here we compare our results from the hybrid nuclear+PP model of the last section with those from solely the PP model containing the four parameters $\log p_1, \Gamma_1, \Gamma_2,\Gamma_3$, where $p_1$ is the pressure at the first dividing density, and $\Gamma_i$ denotes the polytropic indices separated by two fixed transition densities at $1.8 n_0$ and $3.6 n_0$ respectively. We consider similar priors for $\Gamma_i$ as described in Table~\ref{tab:prior} and uniform prior $\in$ (33.5,34.8) for $\log (p_1/\rm {dyn}~ cm^{-2})$. 

\begin{table}[ht]
\begin{tabular}{cccccc} 
\hline
 Quantity& $2.14 M_{\odot}$ pulsar & +GW & +NICER\\

\hline 
\vspace{1ex}

 $R_{1.4} [\rm km]$ &$13.03_{-2.03}^{+4.25}  $ & $11.41_{-0.88}^{+1.56}$ &  $12.26_{-1.24}^{+0.95}$\\

\vspace{1ex}
$\Lambda_{1.4}$ & $713_{-460}^{+1626}$ & $311_{-118}^{+370}$ & $476_{-224}^{+288}$ \\

\vspace{1ex}
$M_{\rm max} (M_{\odot})$ & $2.32_{-0.25}^{+0.37}$ & $2.25_{-0.21}^{+0.34}$ & $2.31_{-0.23}^{+0.43}$ \\

\hline 
\end{tabular}
\caption{Median and $90 \%$ CR of $R_{1.4}$, $\Lambda_{1.4}$ and maximum mass $\rm\ M_{max}$ are quoted here using piecewise polytrope parameterization.   
}
\label{tab-pp}
\end{table}

In Fig.~\ref{fig:EoS-post-pp}, the resulting pressure-density and mass-radius posteriors are plotted in a similar fashion as in Fig.~\ref{fig:post-poly}. By visually inspecting those two figures we can see that the PP parameterization prefers much softer EoS posteriors compared to our hybrid nuclear+PP EoS parameterization. In Table~\ref{tab-pp}, we also report the median and $90 \%$ CR of $\Lambda_{1.4}$, $R_{1.4}$, and $M_{\rm max}$ by adding successive observations using the PP parameterization. With the combined GW and PSR J0740+6620 observations if we compare the lower bound of $R_{1.4}$ coming from these two EoS parameterizations we find that the PP model supports $\sim 0.5$ km lesser value than the hybrid model at $90 \%$ CR. We also observe the similar trend after adding the NICER observation. On the other hand, the upper limit on $R_{1.4}$ coming from these two models are more or less similar after adding the NICER data. Therefore our nuclear physics informed hybrid model helps to rule out a certain softer portion of region in EoS posterior which is allowed by PP parameterization. It can be understood from the prior ranges of these two models also. In the left panel (both upper and lower) of Fig.~\ref{fig:post-poly} and Fig.~\ref{fig:EoS-post-pp}, the corresponding $90 \%$ CR of priors are shown in EoS and mass-radius diagram. We can see that the prior range is much wider for PP in softer region.

Finally, we compute the Bayes factor between two models to find out which one is preferred over the other. In Table~\ref{tab:Bayes-factor}, the values of log-evidences (Z) and Bayes factor (B.F.) between the two models are reported adding successive observations. We follow the Bayes factor significance chart from ~\citet{Bayes-factor} to interpret our result and find no significant evidence for one model over the another.

\begin{table*}[t!]
\begin{tabular}{@{}lllllll@{}}
\toprule
\hline
 Posterior  & \multicolumn{3}{c}{ln(Z)}        & $\ln \rm (BF_{12})$ & $\ln \rm (BF_{13})$ & $\ln \rm (BF_{23})$\\
 \midrule
                        & 1. hybrid NP+PP        & \hspace{5ex}2. PP
                  & 3. NP          \\
\midrule
\hline
+ 2.14 M$_{\odot}$ pulsar & $-1.02$ $\pm$ 0.01  & $-1.68$ $\pm$ 0.01 & $-0.45$ $\pm$ 0.01 &   +0.66 & $-0.57$ & $-1.23$\ \ \ \  \\
+ GW              & $-21.19$ $\pm$ 0.01 & $-21.06$ $\pm$ 0.01 &  $-21.01$ $\pm$ 0.03 & $-0.13$ &  $-0.18$  & $-0.05$  \\
+ NICER                 & $-22.32$ $\pm$ 0.04 & $-22.73$ $\pm$ 0.01 & $-22.11 \pm 0.03 $& +0.51 & $-0.21$ & $-0.62$\\ 
\hline
\bottomrule
\end{tabular}
\caption{Log-evidences (Z) are reported for the three posterior distributions and three parameterizations. Also the log-Bayes' factors $\rm BF_{12}$, $\rm BF_{13}$ and, $\rm BF_{23}$ are computed between hybrid NP+PP and PP, hybrid NP+PP and NP, and PP and NP models respectively. Following the interpretation of \citet{Bayes-factor} there is no significant support for one parameterization over the other. }
\label{tab:Bayes-factor}
\end{table*}

\subsection{Microscopic properties}
\label{micro-prop}
\begin{figure*}[ht]
\centering
\includegraphics[width=\textwidth]{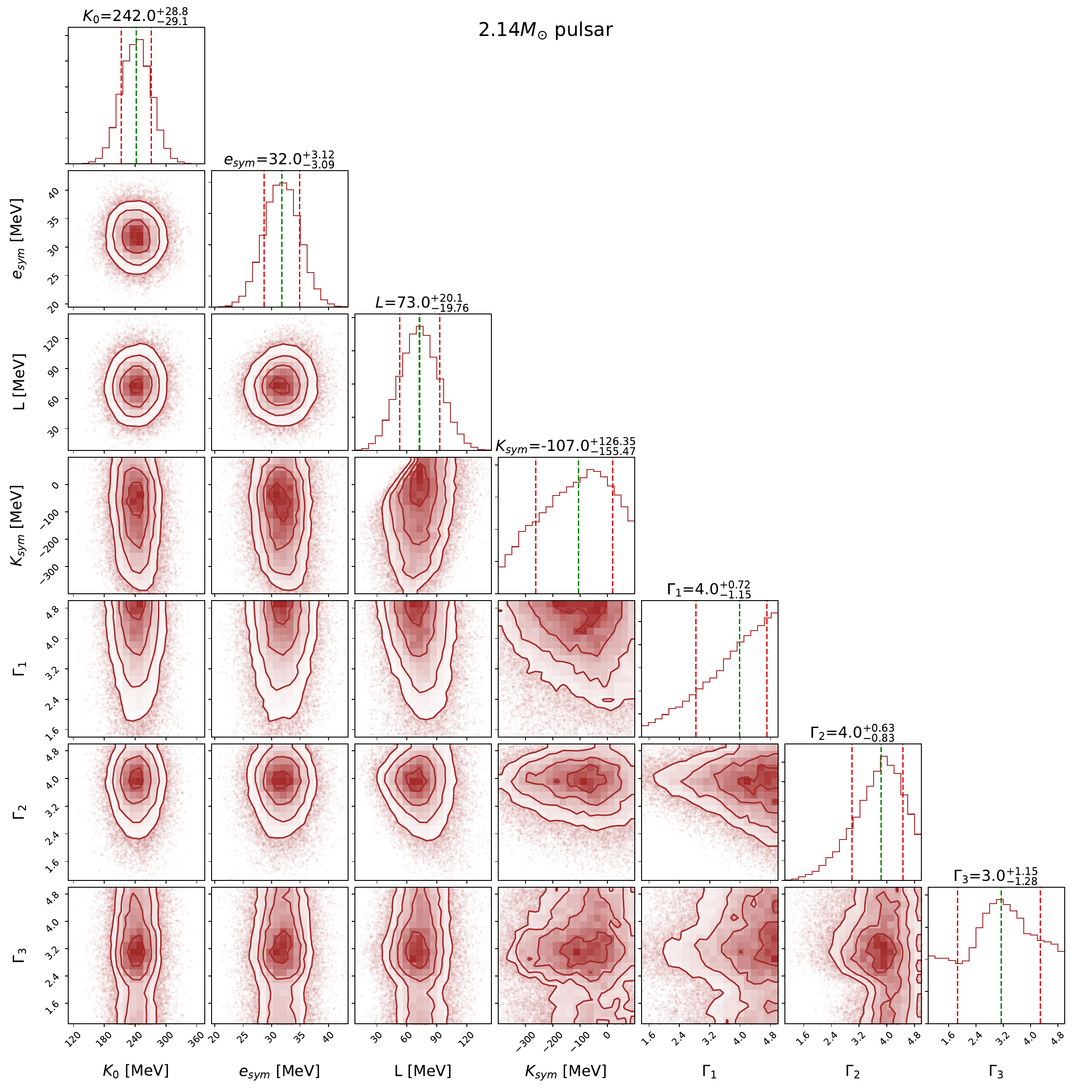}

\caption{Posterior distributions of nuclear parameters after adding PSR J0740+6620 observation are shown using the prior ranges from Table~\ref{tab:prior}.  In the marginalized one-dimensional plots, the median and  $1\sigma$ CR are shown.}
\label{fig:EoS-params-radio}
\end{figure*}

\begin{figure*}[ht]
\centering
\includegraphics[width=\textwidth]{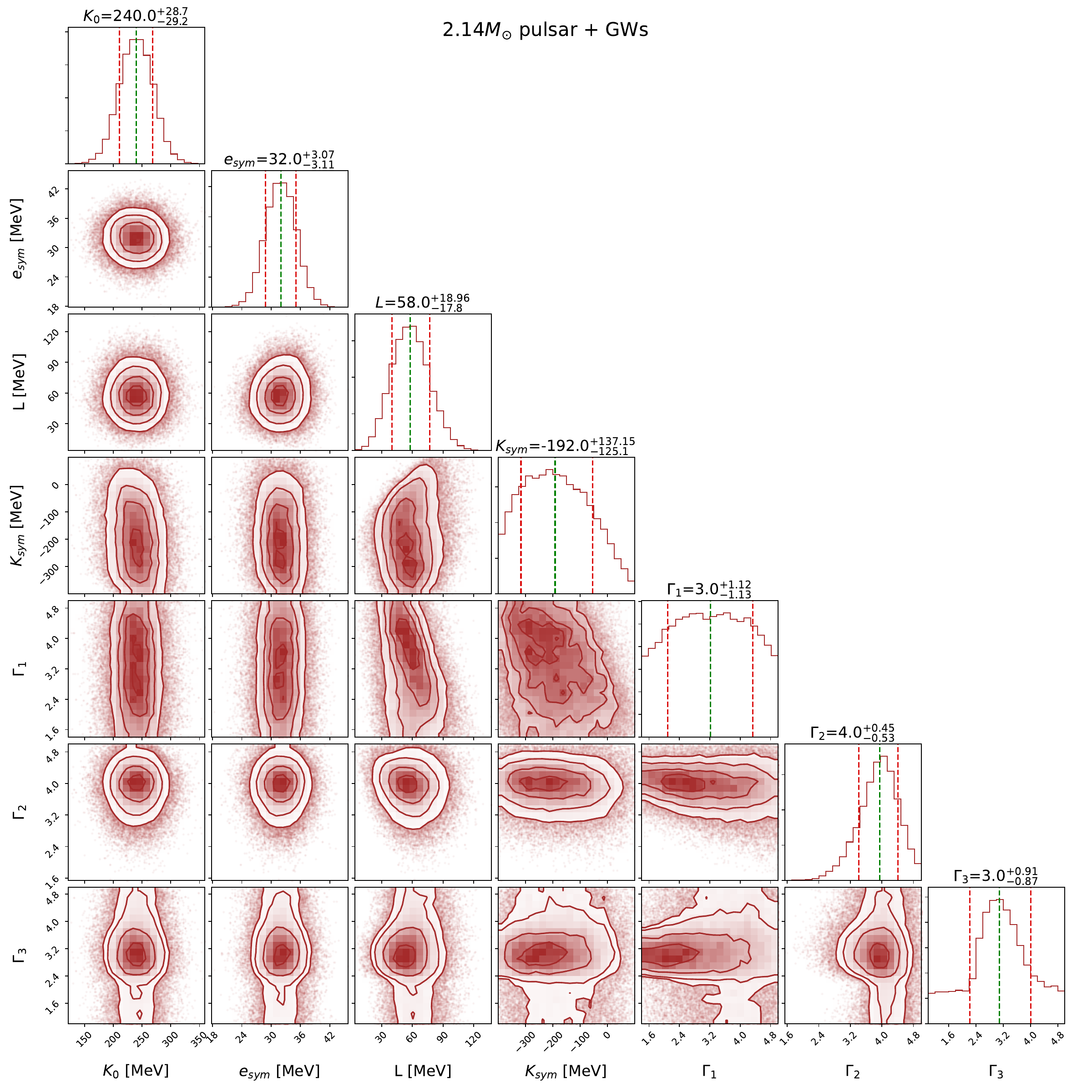}

\caption{Posterior distributions of nuclear parameters after adding PSR J0740+6620 and two GW observations are shown using the prior ranges from Table~\ref{tab:prior}.  In the marginalized one-dimensional plots, the median and  $1\sigma$ CR are shown.}
\label{fig:EoS-params-GW}
\end{figure*}

\begin{figure*}[ht]
\centering
\includegraphics[width=\textwidth]{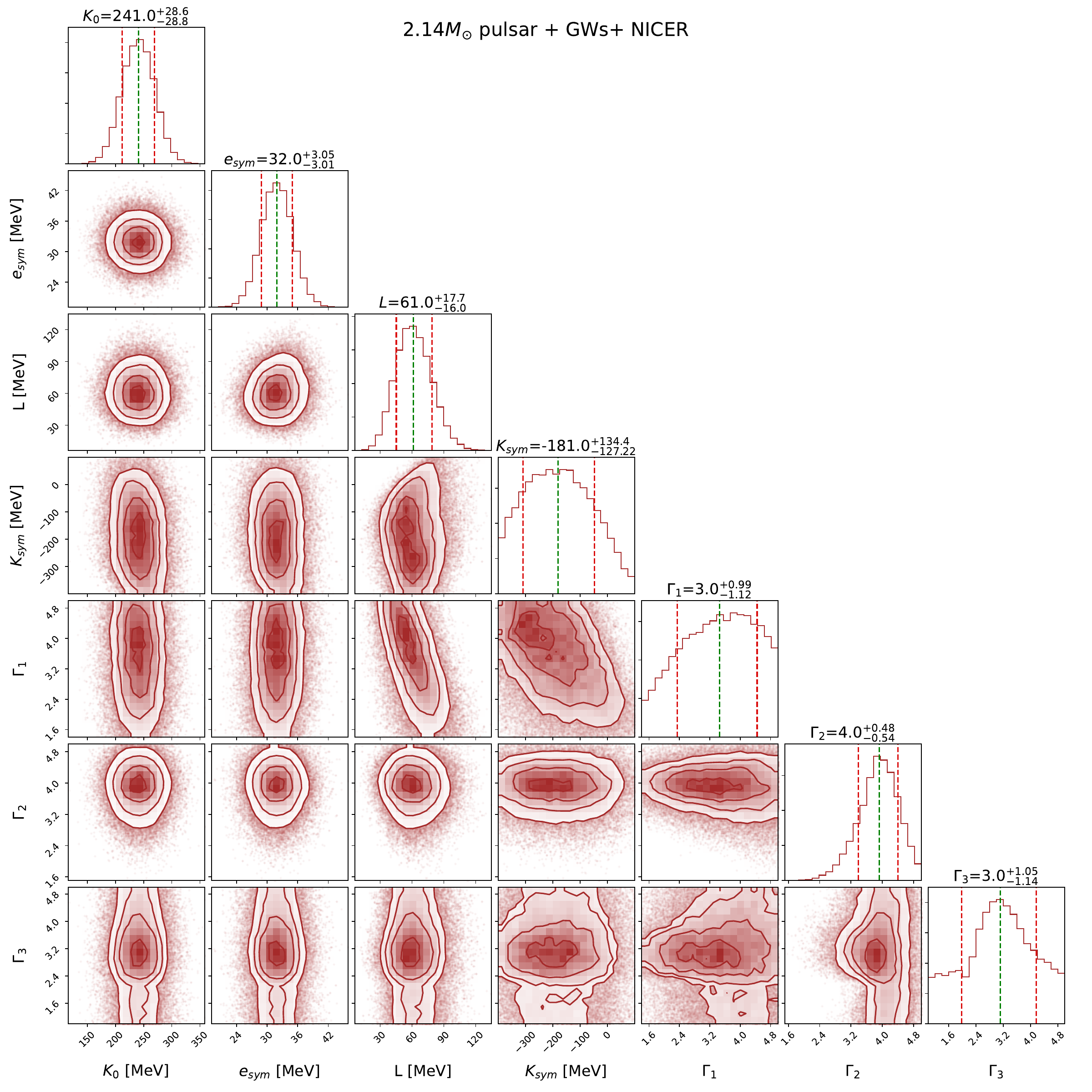}

\caption{Posterior distributions of nuclear parameters after adding PSR J0740+6620, two GW, and NICER observations are shown using the prior ranges from Table~\ref{tab:prior}.  In the marginalized one-dimensional plots, the median and  $1\sigma$ CR are shown.}
\label{fig:EoS-params-GW-NICER}
\end{figure*}

\begin{figure*}[ht]
\centering
\includegraphics[width=\textwidth]{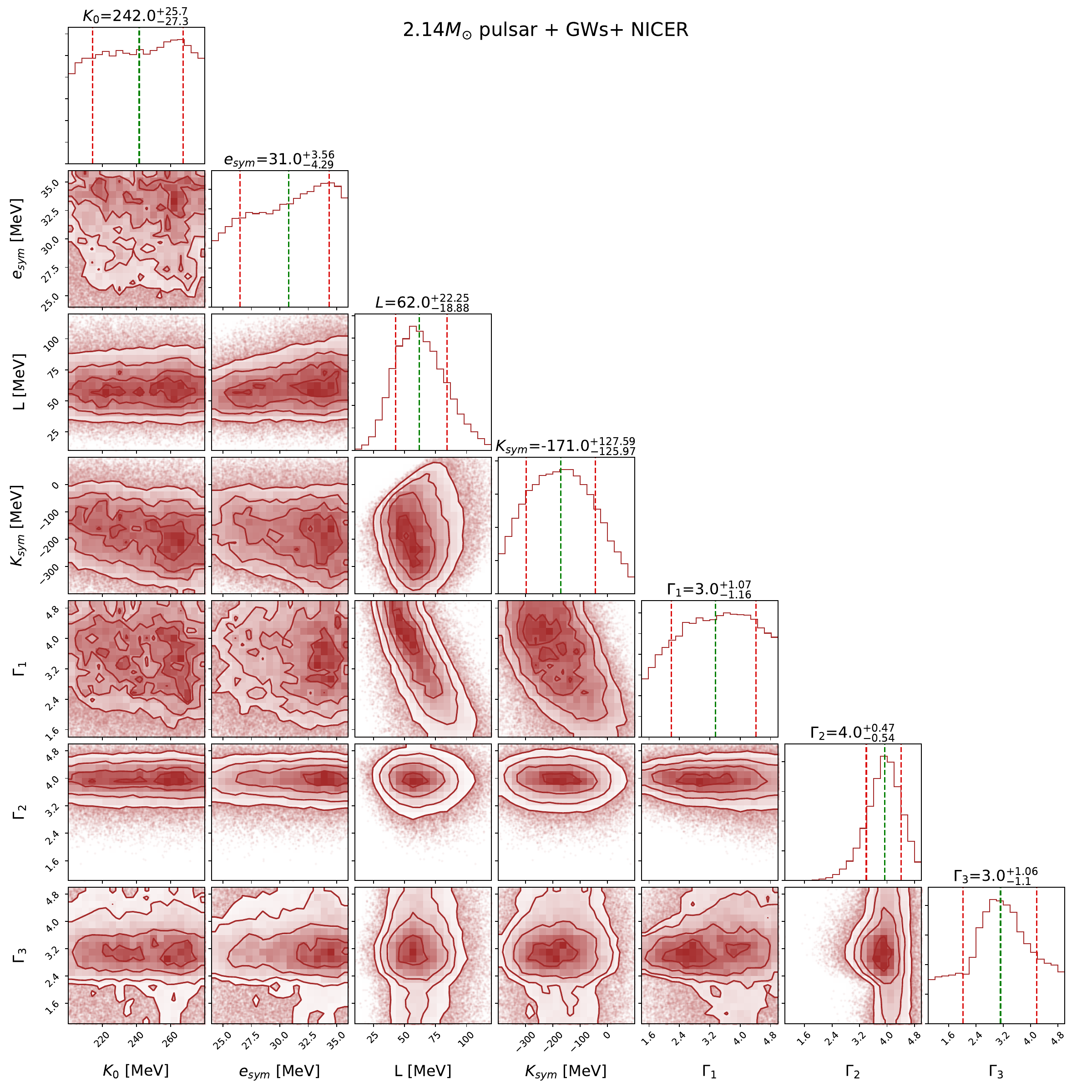}

\caption{This Fig. is same as Fig.~\ref{fig:EoS-params-GW-NICER} but with uniform priors on $K_0$, $e_{\rm sym}$ and $L$.}
\label{fig:EoS-params-uniform-GW-NICER}
\end{figure*}

In ~\Cref{fig:EoS-params-radio,fig:EoS-params-GW,fig:EoS-params-GW-NICER}, the posterior distribution of all the EoS parameters are shown using the prior ranges from Table~\ref{tab:prior} adding successive observations. In the legend of each corner plot the profiles of observational data are indicated. 
For $K_0$ and $e_{\rm sym}$, we do not find much information from the data. We recover the same posterior as the priors for these two parameters. This is not surprising as those are the lowest order nuclear parameters and they have relatively small effect on the macroscopic properties of NS. Given the present statistical error in the macroscopic properties of NS currently we are unable to gain much information about those two lowest order parameters. However we are gaining information about the next higher order parameters $L$ and $K_{\rm sym}$ using the current observational data. Even though we use a strong Gaussian prior of $\mathcal{N} (58.7, 28.1)$ on $L$, using PSR J0740+6620, the bound on $L$ comes out to be $72.6^{+20.1}_{-19.7}$ MeV at $68 \%$ CR. Interestingly, when we combine GW and NICER data its median value again comes very close to the chosen median value of the prior but with a reduced uncertainty. At $68 \%$ CR, we find the bounds on $L$ to be $61.2^{+17.7}_{-16.0}$ MeV. With the combined data, the present constrain on $K_{\rm sym}$ becomes $-181^{+134}_{-127}$ MeV at $68 \%$ CR. In comparison, Ref.~\cite{Zimmerman:2020eho} has obtained a constraint of $K_{\rm sym} =-102^{+71}_{-73}$ MeV. They have used correlations amongst certain combinations of nuclear parameters and radius of $1.4 \rm M_{\odot}$ or tidal deformability based on a very limited number of nuclear EoS. But this type of study is limited by several factors: (1) These correlations highly dependent on how EoS parameters are sampled. In fact, they use a limited number of EoSs, which certainly introduces a bias. To remove this bias proper sampling is required in order to explore all possible ranges in an efficient manner. (2) They didn't include the effect of mass ratio while estimating nuclear parameters. (3) They have studied some particular quantities whereas to get the total picture one has to study a certain minimum number of parameters representative of nuclear EoS in the NS interior. For these reasons they are getting much tighter constraints on $K_{\rm sym}$ than we find. Finally we also explore a scenario [see Figure~\ref{fig:EoS-params-uniform-GW-NICER}] where instead of using a strong Gaussian prior on $K_0$, $e_{\rm sym}$, and $L$, we use uniform priors on these parameters: $K_0 \in (200,280) \rm MeV$, $e_{\rm sym} \in (24,36) \rm MeV$, $L \in (10,120) \rm MeV$. We still find data does not have much power to put any constraint on $K_0$. However, a trend towards the higher $e_{\rm sym}$ can be observed now using the uniform priors. Constraints on $L$, and $K_{\rm sym}$ come closer to the Gaussian prior case, which indicates data is strong enough to put constraints on these two parameters.

\subsection{Apparent tension due to EoS modeling inadequacies}
\label{tension}
\begin{figure*}[ht]
    \centering
    \includegraphics[width=\textwidth]{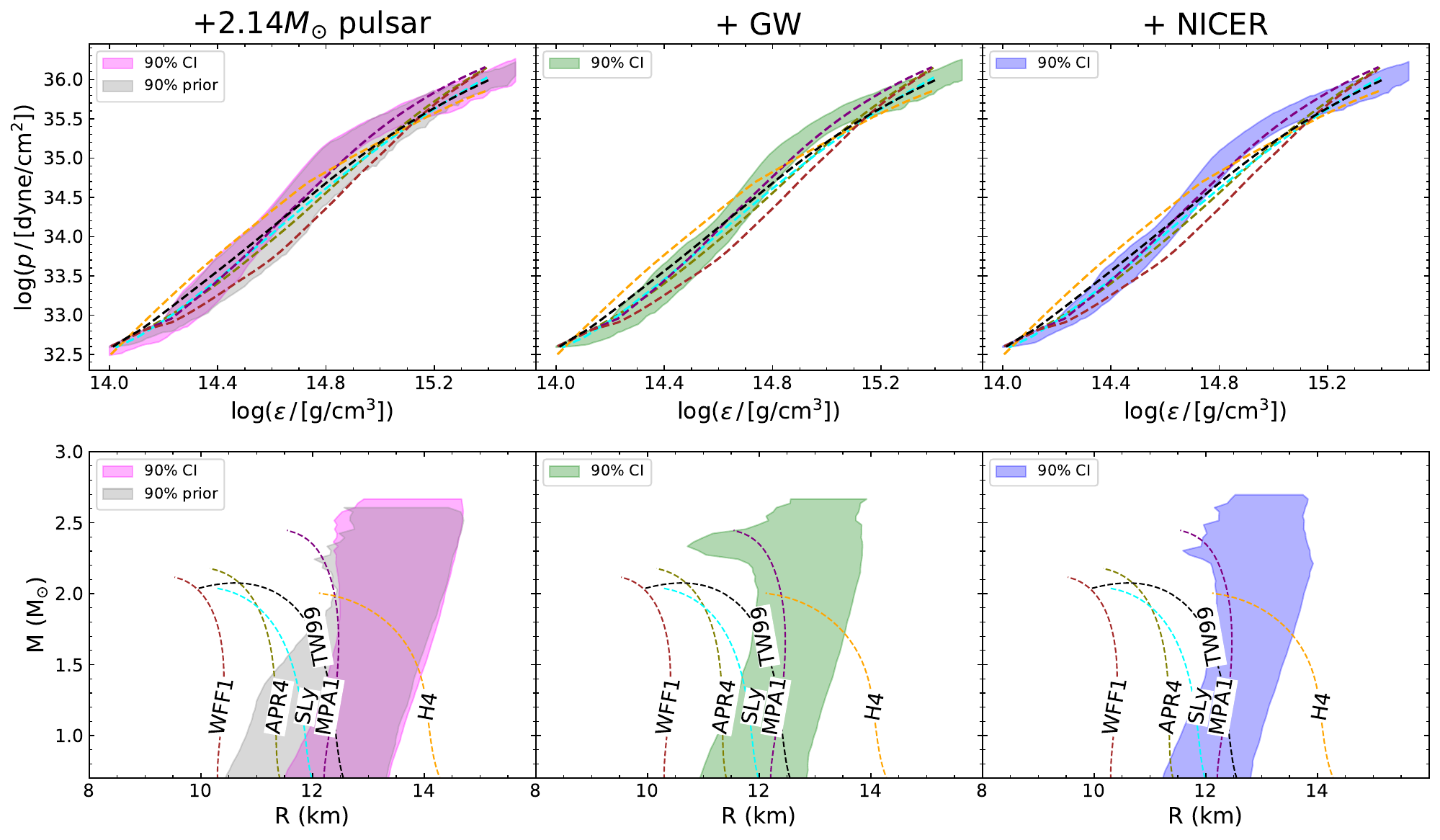}
    \caption{ This figure is similar to Fig.~\ref{fig:post-poly} but with the (inadequate) empirical EoS parameterization.
    }
    \label{fig:Hadronic-post}
\end{figure*}
\begin{table}[ht]
\begin{tabular}{cccccc} 
\hline
 Quantity& $2.14 M_{\odot}$ pulsar & +GW & +NICER\\

\hline 
\vspace{1ex}

 $R_{1.4} [\rm km]$ &$13.24_{-0.98}^{+0.67}  $ & $12.65_{-0.96}^{+0.65}$ &  $12.71_{-0.73}^{+0.58}$\\

\vspace{1ex}
$\Lambda_{1.4}$ & $817_{-316}^{+281}$ & $607_{-238}^{+213}$ & $621_{-196}^{+211}$ \\

\vspace{1ex}
$M_{\rm max} (M_{\odot})$ & $2.39_{-0.28}^{+0.26}$ & $2.34_{-0.28}^{+0.33}$ & $2.35_{-0.28}^{+0.32}$ \\

\hline 
\end{tabular}
\caption{Median and $90 \%$ CR of $R_{1.4}$, $\Lambda_{1.4}$ and maximum mass $\rm\ M_{max}$ are quoted here using nuclear empirical parameterization.   
}
\label{tab-NP}
\end{table}
Previously many authors (see, e.g., Refs.~\cite{Xie:2019sqb,Guven:2020dok,dEtivaux:2019cnf} and the references therein) attempted to constrain NS EoS by employing a parameterization that is based only on the parabolic expansion described in Eq.~\ref{eq:e0} and ~\ref{eq:esym} in Section~\ref{section:EoS}. However, this parameterization is a Taylor expansion around the saturation density,
truncated to 2nd or 3rd order, and then assumed to hold throughout the entire NS. Neutron star cores reach densities multiple times the saturation density (for example, massive stars can reach densities of 5 or 6 times the saturation density). Therefore, at the cores of these stars the expansion parameter is, $\chi = (\rho-\rho_0)/(3\rho_0) \sim (6\rho-\rho_0)/(3\rho_0) \sim 5/3 > 1$. This regime is well outside the region of validity of the Taylor expansion, $\chi \ll 1$. Assuming that the simple Taylor expansion is valid in the entire NS introduces a potentially large and uncontrolled systematic error. To support this assertion, we use this simple, but {\em inadequate}, Taylor-expansion based EoS parameterization~\footnote{ When we extend the empirical parameterization to the high densities, higher order parameters $J_0$, $K_{\rm sym}$, and $J_{\rm sym}$ in Eq.~\ref{eq:e0}, and Eq.~\ref{eq:esym} become more relevant. For this reason instead of second order (as it is done for hybrid NP+PP model) we truncate the the Taylor expansion in $e_0 (\rho_0)$ and $e_{\rm sym} (\rho_0)$ upto third order in $\chi$. we use loose uniform priors on these parameters: $J_0 \in (-800,3000) \rm MeV$, $K_{\rm sym} \in (-800,100) \rm MeV$, $J_{\rm sym} \in (-1000,3000) \rm MeV$. Prior ranges of $e_{\rm sym}$, and $L$ has been kept same as shown in Table~\ref{tab:prior}. } and calculate the corresponding constraint in pressure-density and mass-radius plane using the currently available astrophysical observations. This is shown in Fig.~\ref{fig:Hadronic-post} in a same fashion as the in Fig.~\ref{fig:post-poly} and Fig.~\ref{fig:EoS-post-pp}.
Comparing the posteriors of PP and hybrid nuclear+PP EoS models, we find this simple Taylor expansion based model prefers much stiffer EoS posterior. In Table~\ref{tab-NP}, we report the median and $90 \%$ CR of $R_{1.4}$,$\Lambda_{1.4}$, and $M_{\rm max}$ are quoted adding successive observations using nuclear empirical parameterization. If we compare the lower bound of $R_{1.4}$ with combined GW and PSR J0740+6620 data, we find PP and hybrid NP+PP model supports $\sim 1$ km and $\sim 0.5$ km lesser value than the empirical parameterization model at $90 \%$ CR. Similar trend is also present after adding NICER observation. Therefore, 
there is an apparent tension between the empirical model and the astrophysical data. However, this apparent tension is not conclusively reflected  in the Bayes factors (see Table~\ref{tab:Bayes-factor}), which perhaps indicates the need for more observations with power for distinguishing different EoS models.

Interestingly, the LVC-published EoS-insensitive GW170817 posterior of $\widetilde{\Lambda}$ has a bi-modality: The primary mode peaks at $\widetilde{\Lambda} \sim$ 200 and the secondary at $\widetilde{\Lambda} \sim$ 600. This model tends to prefer the secondary mode, although it is less probable. This results in a posterior preferring somewhat stiffer EoS. However, when we use a better generic parameterization at high densities this apparent tension stands mitigated. This clearly indicates the inconsistency between this particular model, as parameterized here, and the gravitational-wave posteriors of the masses and $\widetilde{\Lambda}$ of GW170817 as given in Ref.~\citet{Abbott:2018exr}.

\section{Conclusion}
\label{summary}

In this work, we adopted a different NS EoS parameterization than what has been used in past studies towards inferring the properties of dense matter from multi-messenger observations. We have made a hybrid parameterization of the EoS by combining two widely used models: nuclear empirical parameters based on the parabolic expansion of binding energy of nucleon around $\rho_0$ and the nuclear-physics agnostic PP model. Based on the Bayesian evidence calculation, we have selected the most likely density where matter properties transition 
from the low-density model employed here to the high-density one. We have shown our results in the context of three different constraints and combinations of them. First, we have selected the constraint arising solely from the observation of massive pulsars (PSR J0740+6620). Irrespective of EoS parameterization, we observe the preferred EoS to be quite stiff. Subsequently, we have added the two GW events (GW170817 \& GW190425), and simultaneous mass-radius measurements of  PSR  J0030+0451 by NICER. Adding only the GW events makes the preferred EoS a lot softer, while further adding NICER makes the EoS stiffer again, although not as much as the sole constraint from pulsar mass observation. Furthermore, the hybrid EoS model constrains $L$ and $K_{\rm sym}$ in a consistent way. In comparison to the purely PP model, the hybrid EoS model prefers stiffer EoSs. But, the combined data do not show any preference for one model over the other. Although, the hybrid model prefers a stiff EoS, we do not find any tension between the nuclear and the astrophysical data. We have discussed this point in detail in section \ref{tension}.
This tension is an artifact that arises when one  applies 
the empirically parameterized EoS to astrophysical data, which shows preference for a different characterization at high densities.

In a future work, we plan to improve our constraints by further adding new data of neutron skin thickness from the PREX experiment \cite{Abrahamyan:2012gp} and study the correlations between the empirical parameters and stellar macroscopic quantities, as was recently done in Ref. \cite{Essick_skin}. In general, our methodology can be extended to include any future observation resulting in stricter constraints on the properties of dense matter.

\section*{Acknowledgments}
We thank the anonymous referees for their useful comments which helped to improve this work significantly. We thank S. Shaikh, N. Paul, V. Prasad, K. S. Phukon, P. Landry, S. Datta and Sudhagar S. for useful discussions. We also thank R. Essick and B. Whiting for carefully reading the manuscript and making several useful suggestions. We gratefully acknowledge the use of high performance super-computing cluster Pegasus at IUCAA for this work. P.C. acknowledges past support from INFN, as a postdoctoral fellow. P.C. is currently supported by the Fonds de la Recherche Scientifique-FNRS, Belgium, under grant No. 4.4503.19.

\bibliography{mybiblio}

\begin{thebibliography}{55}
\expandafter\ifx\csname natexlab\endcsname\relax\def\natexlab#1{#1}\fi
\expandafter\ifx\csname bibnamefont\endcsname\relax
  \def\bibnamefont#1{#1}\fi
\expandafter\ifx\csname bibfnamefont\endcsname\relax
  \def\bibfnamefont#1{#1}\fi
\expandafter\ifx\csname citenamefont\endcsname\relax
  \def\citenamefont#1{#1}\fi
\expandafter\ifx\csname url\endcsname\relax
  \def\url#1{\texttt{#1}}\fi
\expandafter\ifx\csname urlprefix\endcsname\relax\def\urlprefix{URL }\fi
\providecommand{\bibinfo}[2]{#2}
\providecommand{\eprint}[2][]{\url{#2}}

\bibitem[{\citenamefont{Riley et~al.}(2019)}]{Riley:2019yda}
\bibinfo{author}{\bibfnamefont{T.~E.} \bibnamefont{Riley}}
  \bibnamefont{et~al.}, \bibinfo{journal}{Astrophys. J. Lett.}
  \textbf{\bibinfo{volume}{887}}, \bibinfo{pages}{L21} (\bibinfo{year}{2019}),
  \eprint{1912.05702}.

\bibitem[{\citenamefont{Miller et~al.}(2019{\natexlab{a}})}]{Miller:2019cac}
\bibinfo{author}{\bibfnamefont{M.~C.} \bibnamefont{Miller}}
  \bibnamefont{et~al.}, \bibinfo{journal}{Astrophys. J. Lett.}
  \textbf{\bibinfo{volume}{887}}, \bibinfo{pages}{L24}
  (\bibinfo{year}{2019}{\natexlab{a}}), \eprint{1912.05705}.

\bibitem[{\citenamefont{Antoniadis et~al.}(2013)}]{Antoniadis:2013pzd}
\bibinfo{author}{\bibfnamefont{J.}~\bibnamefont{Antoniadis}}
  \bibnamefont{et~al.}, \bibinfo{journal}{Science}
  \textbf{\bibinfo{volume}{340}}, \bibinfo{pages}{6131} (\bibinfo{year}{2013}),
  \eprint{1304.6875}.

\bibitem[{\citenamefont{Cromartie et~al.}(2019)}]{Cromartie:2019kug}
\bibinfo{author}{\bibfnamefont{H.~T.} \bibnamefont{Cromartie}}
  \bibnamefont{et~al.}, \bibinfo{journal}{Nat. Astron.}
  \textbf{\bibinfo{volume}{4}}, \bibinfo{pages}{72} (\bibinfo{year}{2019}),
  \eprint{1904.06759}.

\bibitem[{\citenamefont{Abbott et~al.}(2017)}]{TheLIGOScientific:2017qsa}
\bibinfo{author}{\bibfnamefont{B.~P.} \bibnamefont{Abbott}}
  \bibnamefont{et~al.} (\bibinfo{collaboration}{LIGO Scientific, Virgo}),
  \bibinfo{journal}{Phys. Rev. Lett.} \textbf{\bibinfo{volume}{119}},
  \bibinfo{pages}{161101} (\bibinfo{year}{2017}), \eprint{1710.05832}.

\bibitem[{\citenamefont{Abbott et~al.}(2020)}]{Abbott:2020uma}
\bibinfo{author}{\bibfnamefont{B.~P.} \bibnamefont{Abbott}}
  \bibnamefont{et~al.} (\bibinfo{collaboration}{LIGO Scientific, Virgo})
  (\bibinfo{year}{2020}), \eprint{2001.01761}.

\bibitem[{\citenamefont{Aasi et~al.}(2015)}]{advanced-ligo}
\bibinfo{author}{\bibfnamefont{J.}~\bibnamefont{Aasi}} \bibnamefont{et~al.}
  (\bibinfo{collaboration}{LIGO Scientific}), \bibinfo{journal}{Class. Quant.
  Grav.} \textbf{\bibinfo{volume}{32}}, \bibinfo{pages}{074001}
  (\bibinfo{year}{2015}), \eprint{1411.4547}.

\bibitem[{\citenamefont{Acernese et~al.}(2015)}]{advanced-virgo}
\bibinfo{author}{\bibfnamefont{F.}~\bibnamefont{Acernese}} \bibnamefont{et~al.}
  (\bibinfo{collaboration}{VIRGO}), \bibinfo{journal}{Class. Quant. Grav.}
  \textbf{\bibinfo{volume}{32}}, \bibinfo{pages}{024001}
  (\bibinfo{year}{2015}), \eprint{1408.3978}.

\bibitem[{\citenamefont{Abbott et~al.}(2018)}]{Abbott:2018exr}
\bibinfo{author}{\bibfnamefont{B.~P.} \bibnamefont{Abbott}}
  \bibnamefont{et~al.} (\bibinfo{collaboration}{LIGO Scientific, Virgo}),
  \bibinfo{journal}{Phys. Rev. Lett.} \textbf{\bibinfo{volume}{121}},
  \bibinfo{pages}{161101} (\bibinfo{year}{2018}), \eprint{1805.11581}.

\bibitem[{\citenamefont{Raaijmakers et~al.}(2020)}]{Raaijmakers:2019dks}
\bibinfo{author}{\bibfnamefont{G.}~\bibnamefont{Raaijmakers}}
  \bibnamefont{et~al.}, \bibinfo{journal}{Astrophys. J. Lett.}
  \textbf{\bibinfo{volume}{893}}, \bibinfo{pages}{L21} (\bibinfo{year}{2020}),
  \eprint{1912.11031}.

\bibitem[{\citenamefont{Jiang et~al.}(2020)\citenamefont{Jiang, Tang, Wang,
  Fan, and Wei}}]{Jiang:2019rcw}
\bibinfo{author}{\bibfnamefont{J.-L.} \bibnamefont{Jiang}},
  \bibinfo{author}{\bibfnamefont{S.-P.} \bibnamefont{Tang}},
  \bibinfo{author}{\bibfnamefont{Y.-Z.} \bibnamefont{Wang}},
  \bibinfo{author}{\bibfnamefont{Y.-Z.} \bibnamefont{Fan}}, \bibnamefont{and}
  \bibinfo{author}{\bibfnamefont{D.-M.} \bibnamefont{Wei}},
  \bibinfo{journal}{Astrophys. J.} \textbf{\bibinfo{volume}{892}},
  \bibinfo{pages}{1} (\bibinfo{year}{2020}), \eprint{1912.07467}.

\bibitem[{\citenamefont{Landry et~al.}(2020)\citenamefont{Landry, Essick, and
  Chatziioannou}}]{Landry:2020vaw}
\bibinfo{author}{\bibfnamefont{P.}~\bibnamefont{Landry}},
  \bibinfo{author}{\bibfnamefont{R.}~\bibnamefont{Essick}}, \bibnamefont{and}
  \bibinfo{author}{\bibfnamefont{K.}~\bibnamefont{Chatziioannou}}
  (\bibinfo{year}{2020}), \eprint{2003.04880}.

\bibitem[{\citenamefont{Read et~al.}(2009)\citenamefont{Read, Lackey, Owen, and
  Friedman}}]{Read:2008iy}
\bibinfo{author}{\bibfnamefont{J.~S.} \bibnamefont{Read}},
  \bibinfo{author}{\bibfnamefont{B.~D.} \bibnamefont{Lackey}},
  \bibinfo{author}{\bibfnamefont{B.~J.} \bibnamefont{Owen}}, \bibnamefont{and}
  \bibinfo{author}{\bibfnamefont{J.~L.} \bibnamefont{Friedman}},
  \bibinfo{journal}{Phys. Rev.} \textbf{\bibinfo{volume}{D79}},
  \bibinfo{pages}{124032} (\bibinfo{year}{2009}), \eprint{0812.2163}.

\bibitem[{\citenamefont{Lindblom}(2010)}]{Lindblom:2010bb}
\bibinfo{author}{\bibfnamefont{L.}~\bibnamefont{Lindblom}},
  \bibinfo{journal}{Phys. Rev.} \textbf{\bibinfo{volume}{D82}},
  \bibinfo{pages}{103011} (\bibinfo{year}{2010}), \eprint{1009.0738}.

\bibitem[{\citenamefont{{Lackey} and {Wade}}(2015)}]{Lackey-2015}
\bibinfo{author}{\bibfnamefont{B.~D.} \bibnamefont{{Lackey}}} \bibnamefont{and}
  \bibinfo{author}{\bibfnamefont{L.}~\bibnamefont{{Wade}}},
  \bibinfo{journal}{\prd} \textbf{\bibinfo{volume}{91}}, \bibinfo{eid}{043002}
  (\bibinfo{year}{2015}), \eprint{1410.8866}.

\bibitem[{\citenamefont{Carney et~al.}(2018)\citenamefont{Carney, Wade, and
  Irwin}}]{Carney:2018sdv}
\bibinfo{author}{\bibfnamefont{M.~F.} \bibnamefont{Carney}},
  \bibinfo{author}{\bibfnamefont{L.~E.} \bibnamefont{Wade}}, \bibnamefont{and}
  \bibinfo{author}{\bibfnamefont{B.~S.} \bibnamefont{Irwin}},
  \bibinfo{journal}{Phys. Rev. D} \textbf{\bibinfo{volume}{98}},
  \bibinfo{pages}{063004} (\bibinfo{year}{2018}), \eprint{1805.11217}.

\bibitem[{\citenamefont{Landry and Essick}(2019)}]{Landry:2018prl}
\bibinfo{author}{\bibfnamefont{P.}~\bibnamefont{Landry}} \bibnamefont{and}
  \bibinfo{author}{\bibfnamefont{R.}~\bibnamefont{Essick}},
  \bibinfo{journal}{Phys. Rev. D} \textbf{\bibinfo{volume}{99}},
  \bibinfo{pages}{084049} (\bibinfo{year}{2019}), \eprint{1811.12529}.

\bibitem[{\citenamefont{Essick et~al.}(2019)\citenamefont{Essick, Landry, and
  Holz}}]{Essick:2019ldf}
\bibinfo{author}{\bibfnamefont{R.}~\bibnamefont{Essick}},
  \bibinfo{author}{\bibfnamefont{P.}~\bibnamefont{Landry}}, \bibnamefont{and}
  \bibinfo{author}{\bibfnamefont{D.~E.} \bibnamefont{Holz}}
  (\bibinfo{year}{2019}), \eprint{1910.09740}.

\bibitem[{\citenamefont{Capano et~al.}(2020)\citenamefont{Capano, Tews, Brown,
  Margalit, De, Kumar, Brown, Krishnan, and Reddy}}]{Capano:2019eae}
\bibinfo{author}{\bibfnamefont{C.~D.} \bibnamefont{Capano}},
  \bibinfo{author}{\bibfnamefont{I.}~\bibnamefont{Tews}},
  \bibinfo{author}{\bibfnamefont{S.~M.} \bibnamefont{Brown}},
  \bibinfo{author}{\bibfnamefont{B.}~\bibnamefont{Margalit}},
  \bibinfo{author}{\bibfnamefont{S.}~\bibnamefont{De}},
  \bibinfo{author}{\bibfnamefont{S.}~\bibnamefont{Kumar}},
  \bibinfo{author}{\bibfnamefont{D.~A.} \bibnamefont{Brown}},
  \bibinfo{author}{\bibfnamefont{B.}~\bibnamefont{Krishnan}}, \bibnamefont{and}
  \bibinfo{author}{\bibfnamefont{S.}~\bibnamefont{Reddy}},
  \bibinfo{journal}{Nature Astron.} \textbf{\bibinfo{volume}{4}},
  \bibinfo{pages}{625} (\bibinfo{year}{2020}), \eprint{1908.10352}.

\bibitem[{\citenamefont{{Essick} et~al.}(2020)\citenamefont{{Essick}, {Tews},
  {Landry}, {Reddy}, and {Holz}}}]{Essick-2020arXiv}
\bibinfo{author}{\bibfnamefont{R.}~\bibnamefont{{Essick}}},
  \bibinfo{author}{\bibfnamefont{I.}~\bibnamefont{{Tews}}},
  \bibinfo{author}{\bibfnamefont{P.}~\bibnamefont{{Landry}}},
  \bibinfo{author}{\bibfnamefont{S.}~\bibnamefont{{Reddy}}}, \bibnamefont{and}
  \bibinfo{author}{\bibfnamefont{D.~E.} \bibnamefont{{Holz}}},
  \bibinfo{journal}{arXiv e-prints} \bibinfo{eid}{arXiv:2004.07744}
  (\bibinfo{year}{2020}), \eprint{2004.07744}.

\bibitem[{\citenamefont{Piekarewicz and Centelles}(2009)}]{Piekarewicz:2008nh}
\bibinfo{author}{\bibfnamefont{J.}~\bibnamefont{Piekarewicz}} \bibnamefont{and}
  \bibinfo{author}{\bibfnamefont{M.}~\bibnamefont{Centelles}},
  \bibinfo{journal}{Phys. Rev. C} \textbf{\bibinfo{volume}{79}},
  \bibinfo{pages}{054311} (\bibinfo{year}{2009}), \eprint{0812.4499}.

\bibitem[{\citenamefont{Steiner et~al.}(2010)\citenamefont{Steiner, Lattimer,
  and Brown}}]{Steiner:2010fz}
\bibinfo{author}{\bibfnamefont{A.~W.} \bibnamefont{Steiner}},
  \bibinfo{author}{\bibfnamefont{J.~M.} \bibnamefont{Lattimer}},
  \bibnamefont{and} \bibinfo{author}{\bibfnamefont{E.~F.} \bibnamefont{Brown}},
  \bibinfo{journal}{Astrophys. J.} \textbf{\bibinfo{volume}{722}},
  \bibinfo{pages}{33} (\bibinfo{year}{2010}), \eprint{1005.0811}.

\bibitem[{\citenamefont{Güven et~al.}(2020)\citenamefont{Güven, Bozkurt,
  Khan, and Margueron}}]{Guven:2020dok}
\bibinfo{author}{\bibfnamefont{H.}~\bibnamefont{Güven}},
  \bibinfo{author}{\bibfnamefont{K.}~\bibnamefont{Bozkurt}},
  \bibinfo{author}{\bibfnamefont{E.}~\bibnamefont{Khan}}, \bibnamefont{and}
  \bibinfo{author}{\bibfnamefont{J.}~\bibnamefont{Margueron}}
  (\bibinfo{year}{2020}), \eprint{2001.10259}.

\bibitem[{\citenamefont{Zhang and Li}(2019)}]{Zhang:2019fog}
\bibinfo{author}{\bibfnamefont{N.-B.} \bibnamefont{Zhang}} \bibnamefont{and}
  \bibinfo{author}{\bibfnamefont{B.-A.} \bibnamefont{Li}},
  \bibinfo{journal}{Astrophys. J.} \textbf{\bibinfo{volume}{879}},
  \bibinfo{pages}{99} (\bibinfo{year}{2019}), \eprint{1904.10998}.

\bibitem[{\citenamefont{Xie and Li}(2019)}]{Xie:2019sqb}
\bibinfo{author}{\bibfnamefont{W.-J.} \bibnamefont{Xie}} \bibnamefont{and}
  \bibinfo{author}{\bibfnamefont{B.-A.} \bibnamefont{Li}},
  \bibinfo{journal}{Astrophys. J.} \textbf{\bibinfo{volume}{883}},
  \bibinfo{pages}{174} (\bibinfo{year}{2019}), \eprint{1907.10741}.

\bibitem[{\citenamefont{Baillot~d'Etivaux
  et~al.}(2019)\citenamefont{Baillot~d'Etivaux, Guillot, Margueron, Webb,
  Catelan, and Reisenegger}}]{dEtivaux:2019cnf}
\bibinfo{author}{\bibfnamefont{N.}~\bibnamefont{Baillot~d'Etivaux}},
  \bibinfo{author}{\bibfnamefont{S.}~\bibnamefont{Guillot}},
  \bibinfo{author}{\bibfnamefont{J.}~\bibnamefont{Margueron}},
  \bibinfo{author}{\bibfnamefont{N.}~\bibnamefont{Webb}},
  \bibinfo{author}{\bibfnamefont{M.}~\bibnamefont{Catelan}}, \bibnamefont{and}
  \bibinfo{author}{\bibfnamefont{A.}~\bibnamefont{Reisenegger}}
  (\bibinfo{year}{2019}), \eprint{1905.01081}.

\bibitem[{\citenamefont{{Carson} et~al.}(2019)\citenamefont{{Carson},
  {Steiner}, and {Yagi}}}]{Carson2019:nucl170817}
\bibinfo{author}{\bibfnamefont{Z.}~\bibnamefont{{Carson}}},
  \bibinfo{author}{\bibfnamefont{A.~W.} \bibnamefont{{Steiner}}},
  \bibnamefont{and} \bibinfo{author}{\bibfnamefont{K.}~\bibnamefont{{Yagi}}},
  \bibinfo{journal}{\prd} \textbf{\bibinfo{volume}{99}}, \bibinfo{eid}{043010}
  (\bibinfo{year}{2019}), \eprint{1812.08910}.

\bibitem[{\citenamefont{Zimmerman et~al.}(2020)\citenamefont{Zimmerman, Carson,
  Schumacher, Steiner, and Yagi}}]{Zimmerman:2020eho}
\bibinfo{author}{\bibfnamefont{J.}~\bibnamefont{Zimmerman}},
  \bibinfo{author}{\bibfnamefont{Z.}~\bibnamefont{Carson}},
  \bibinfo{author}{\bibfnamefont{K.}~\bibnamefont{Schumacher}},
  \bibinfo{author}{\bibfnamefont{A.~W.} \bibnamefont{Steiner}},
  \bibnamefont{and} \bibinfo{author}{\bibfnamefont{K.}~\bibnamefont{Yagi}}
  (\bibinfo{year}{2020}), \eprint{2002.03210}.

\bibitem[{\citenamefont{Xie and Li}(2020)}]{Xie:2020tdo}
\bibinfo{author}{\bibfnamefont{W.-J.} \bibnamefont{Xie}} \bibnamefont{and}
  \bibinfo{author}{\bibfnamefont{B.-A.} \bibnamefont{Li}}
  (\bibinfo{year}{2020}), \eprint{2005.07216}.

\bibitem[{\citenamefont{Skilling}(2004)}]{Skilling_Nested}
\bibinfo{author}{\bibfnamefont{J.}~\bibnamefont{Skilling}},
  \bibinfo{journal}{AIP Conference Proceedings} \textbf{\bibinfo{volume}{735}},
  \bibinfo{pages}{395} (\bibinfo{year}{2004}),
  \eprint{https://aip.scitation.org/doi/pdf/10.1063/1.1835238},
  \urlprefix\url{https://aip.scitation.org/doi/abs/10.1063/1.1835238}.

\bibitem[{\citenamefont{Margueron et~al.}(2018)\citenamefont{Margueron,
  Hoffmann~Casali, and Gulminelli}}]{Margueron:2017eqc}
\bibinfo{author}{\bibfnamefont{J.}~\bibnamefont{Margueron}},
  \bibinfo{author}{\bibfnamefont{R.}~\bibnamefont{Hoffmann~Casali}},
  \bibnamefont{and}
  \bibinfo{author}{\bibfnamefont{F.}~\bibnamefont{Gulminelli}},
  \bibinfo{journal}{Phys. Rev.} \textbf{\bibinfo{volume}{C97}},
  \bibinfo{pages}{025805} (\bibinfo{year}{2018}), \eprint{1708.06894}.

\bibitem[{\citenamefont{Brown and Schwenk}(2014)}]{Brown:2013pwa}
\bibinfo{author}{\bibfnamefont{B.~A.} \bibnamefont{Brown}} \bibnamefont{and}
  \bibinfo{author}{\bibfnamefont{A.}~\bibnamefont{Schwenk}},
  \bibinfo{journal}{Phys. Rev. C} \textbf{\bibinfo{volume}{89}},
  \bibinfo{pages}{011307} (\bibinfo{year}{2014}), \bibinfo{note}{[Erratum:
  Phys.Rev.C 91, 049902 (2015)]}, \eprint{1311.3957}.

\bibitem[{\citenamefont{Tsang et~al.}(2019)\citenamefont{Tsang, Brown,
  Fattoyev, Lynch, and Tsang}}]{Tsang:2019ymt}
\bibinfo{author}{\bibfnamefont{C.}~\bibnamefont{Tsang}},
  \bibinfo{author}{\bibfnamefont{B.}~\bibnamefont{Brown}},
  \bibinfo{author}{\bibfnamefont{F.}~\bibnamefont{Fattoyev}},
  \bibinfo{author}{\bibfnamefont{W.}~\bibnamefont{Lynch}}, \bibnamefont{and}
  \bibinfo{author}{\bibfnamefont{M.}~\bibnamefont{Tsang}},
  \bibinfo{journal}{Phys. Rev. C} \textbf{\bibinfo{volume}{100}},
  \bibinfo{pages}{062801} (\bibinfo{year}{2019}), \eprint{1908.11842}.

\bibitem[{\citenamefont{Piekarewicz}(2010)}]{Piekarewicz:2009gb}
\bibinfo{author}{\bibfnamefont{J.}~\bibnamefont{Piekarewicz}},
  \bibinfo{journal}{J. Phys. G} \textbf{\bibinfo{volume}{37}},
  \bibinfo{pages}{064038} (\bibinfo{year}{2010}), \eprint{0912.5103}.

\bibitem[{\citenamefont{Lattimer and Lim}(2013)}]{Lattimer:2012xj}
\bibinfo{author}{\bibfnamefont{J.~M.} \bibnamefont{Lattimer}} \bibnamefont{and}
  \bibinfo{author}{\bibfnamefont{Y.}~\bibnamefont{Lim}},
  \bibinfo{journal}{Astrophys. J.} \textbf{\bibinfo{volume}{771}},
  \bibinfo{pages}{51} (\bibinfo{year}{2013}), \eprint{1203.4286}.

\bibitem[{\citenamefont{Oertel et~al.}(2017)\citenamefont{Oertel, Hempel,
  Klähn, and Typel}}]{Oertel:2016bki}
\bibinfo{author}{\bibfnamefont{M.}~\bibnamefont{Oertel}},
  \bibinfo{author}{\bibfnamefont{M.}~\bibnamefont{Hempel}},
  \bibinfo{author}{\bibfnamefont{T.}~\bibnamefont{Klähn}}, \bibnamefont{and}
  \bibinfo{author}{\bibfnamefont{S.}~\bibnamefont{Typel}},
  \bibinfo{journal}{Rev. Mod. Phys.} \textbf{\bibinfo{volume}{89}},
  \bibinfo{pages}{015007} (\bibinfo{year}{2017}), \eprint{1610.03361}.

\bibitem[{\citenamefont{Tews et~al.}(2017)\citenamefont{Tews, Lattimer,
  Ohnishi, and Kolomeitsev}}]{Kolomeitsev:2016sjl}
\bibinfo{author}{\bibfnamefont{I.}~\bibnamefont{Tews}},
  \bibinfo{author}{\bibfnamefont{J.~M.} \bibnamefont{Lattimer}},
  \bibinfo{author}{\bibfnamefont{A.}~\bibnamefont{Ohnishi}}, \bibnamefont{and}
  \bibinfo{author}{\bibfnamefont{E.~E.} \bibnamefont{Kolomeitsev}},
  \bibinfo{journal}{Astrophys. J.} \textbf{\bibinfo{volume}{848}},
  \bibinfo{pages}{105} (\bibinfo{year}{2017}), \eprint{1611.07133}.

\bibitem[{\citenamefont{Margueron and Gulminelli}(2019)}]{Margueron:2018eob}
\bibinfo{author}{\bibfnamefont{J.}~\bibnamefont{Margueron}} \bibnamefont{and}
  \bibinfo{author}{\bibfnamefont{F.}~\bibnamefont{Gulminelli}},
  \bibinfo{journal}{Phys. Rev. C} \textbf{\bibinfo{volume}{99}},
  \bibinfo{pages}{025806} (\bibinfo{year}{2019}), \eprint{1807.01729}.

\bibitem[{\citenamefont{Kass and Raftery}(1995)}]{Bayes-factor}
\bibinfo{author}{\bibfnamefont{R.~E.} \bibnamefont{Kass}} \bibnamefont{and}
  \bibinfo{author}{\bibfnamefont{A.~E.} \bibnamefont{Raftery}},
  \bibinfo{journal}{Journal of the American Statistical Association}
  \textbf{\bibinfo{volume}{90}}, \bibinfo{pages}{773} (\bibinfo{year}{1995}),
  \eprint{https://amstat.tandfonline.com/doi/pdf/10.1080/01621459.1995.10476572},
  \urlprefix\url{https://amstat.tandfonline.com/doi/abs/10.1080/01621459.1995.10476572}.

\bibitem[{\citenamefont{Biswas et~al.}(2019)\citenamefont{Biswas, Nandi, Char,
  and Bose}}]{Biswas:2019ifs}
\bibinfo{author}{\bibfnamefont{B.}~\bibnamefont{Biswas}},
  \bibinfo{author}{\bibfnamefont{R.}~\bibnamefont{Nandi}},
  \bibinfo{author}{\bibfnamefont{P.}~\bibnamefont{Char}}, \bibnamefont{and}
  \bibinfo{author}{\bibfnamefont{S.}~\bibnamefont{Bose}},
  \bibinfo{journal}{Phys. Rev. D} \textbf{\bibinfo{volume}{100}},
  \bibinfo{pages}{044056} (\bibinfo{year}{2019}), \eprint{1905.00678}.

\bibitem[{\citenamefont{Gamba et~al.}(2020)\citenamefont{Gamba, Read, and
  Wade}}]{Gamba:2019kwu}
\bibinfo{author}{\bibfnamefont{R.}~\bibnamefont{Gamba}},
  \bibinfo{author}{\bibfnamefont{J.~S.} \bibnamefont{Read}}, \bibnamefont{and}
  \bibinfo{author}{\bibfnamefont{L.~E.} \bibnamefont{Wade}},
  \bibinfo{journal}{Class. Quant. Grav.} \textbf{\bibinfo{volume}{37}},
  \bibinfo{pages}{025008} (\bibinfo{year}{2020}), \eprint{1902.04616}.

\bibitem[{\citenamefont{{Baym} et~al.}(1971)\citenamefont{{Baym}, {Pethick},
  and {Sutherland}}}]{1971ApJ...170..299B}
\bibinfo{author}{\bibfnamefont{G.}~\bibnamefont{{Baym}}},
  \bibinfo{author}{\bibfnamefont{C.}~\bibnamefont{{Pethick}}},
  \bibnamefont{and}
  \bibinfo{author}{\bibfnamefont{P.}~\bibnamefont{{Sutherland}}},
  \bibinfo{journal}{\apj} \textbf{\bibinfo{volume}{170}}, \bibinfo{pages}{299}
  (\bibinfo{year}{1971}).

\bibitem[{\citenamefont{Miller et~al.}(2019{\natexlab{b}})\citenamefont{Miller,
  Chirenti, and Lamb}}]{Miller:2019nzo}
\bibinfo{author}{\bibfnamefont{M.~C.} \bibnamefont{Miller}},
  \bibinfo{author}{\bibfnamefont{C.}~\bibnamefont{Chirenti}}, \bibnamefont{and}
  \bibinfo{author}{\bibfnamefont{F.~K.} \bibnamefont{Lamb}}
  (\bibinfo{year}{2019}{\natexlab{b}}), \eprint{1904.08907}.

\bibitem[{\citenamefont{Buchner et~al.}(2014)\citenamefont{Buchner,
  Georgakakis, Nandra, Hsu, Rangel, Brightman, Merloni, Salvato, Donley, and
  Kocevski}}]{pymultinest}
\bibinfo{author}{\bibfnamefont{J.}~\bibnamefont{Buchner}},
  \bibinfo{author}{\bibfnamefont{A.}~\bibnamefont{Georgakakis}},
  \bibinfo{author}{\bibfnamefont{K.}~\bibnamefont{Nandra}},
  \bibinfo{author}{\bibfnamefont{L.}~\bibnamefont{Hsu}},
  \bibinfo{author}{\bibfnamefont{C.}~\bibnamefont{Rangel}},
  \bibinfo{author}{\bibfnamefont{M.}~\bibnamefont{Brightman}},
  \bibinfo{author}{\bibfnamefont{A.}~\bibnamefont{Merloni}},
  \bibinfo{author}{\bibfnamefont{M.}~\bibnamefont{Salvato}},
  \bibinfo{author}{\bibfnamefont{J.}~\bibnamefont{Donley}}, \bibnamefont{and}
  \bibinfo{author}{\bibfnamefont{D.}~\bibnamefont{Kocevski}},
  \bibinfo{journal}{Astron. Astrophys.} \textbf{\bibinfo{volume}{564}},
  \bibinfo{pages}{A125} (\bibinfo{year}{2014}), \eprint{1402.0004}.

\bibitem[{\citenamefont{{Wysocki} et~al.}(2020)\citenamefont{{Wysocki},
  {O'Shaughnessy}, {Wade}, and {Lange}}}]{Wysocki-2020}
\bibinfo{author}{\bibfnamefont{D.}~\bibnamefont{{Wysocki}}},
  \bibinfo{author}{\bibfnamefont{R.}~\bibnamefont{{O'Shaughnessy}}},
  \bibinfo{author}{\bibfnamefont{L.}~\bibnamefont{{Wade}}}, \bibnamefont{and}
  \bibinfo{author}{\bibfnamefont{J.}~\bibnamefont{{Lange}}},
  \bibinfo{journal}{arXiv e-prints} \bibinfo{eid}{arXiv:2001.01747}
  (\bibinfo{year}{2020}), \eprint{2001.01747}.

\bibitem[{\citenamefont{{Akmal} et~al.}(1998)\citenamefont{{Akmal},
  {Pandharipande}, and {Ravenhall}}}]{1998PhRvC..58.1804A}
\bibinfo{author}{\bibfnamefont{A.}~\bibnamefont{{Akmal}}},
  \bibinfo{author}{\bibfnamefont{V.~R.} \bibnamefont{{Pandharipande}}},
  \bibnamefont{and} \bibinfo{author}{\bibfnamefont{D.~G.}
  \bibnamefont{{Ravenhall}}}, \bibinfo{journal}{\prc}
  \textbf{\bibinfo{volume}{58}}, \bibinfo{pages}{1804} (\bibinfo{year}{1998}),
  \eprint{nucl-th/9804027}.

\bibitem[{\citenamefont{Douchin and Haensel}(2001)}]{2001A&A...380..151D}
\bibinfo{author}{\bibfnamefont{F.}~\bibnamefont{Douchin}} \bibnamefont{and}
  \bibinfo{author}{\bibfnamefont{P.}~\bibnamefont{Haensel}},
  \bibinfo{journal}{Astron. Astrophys.} \textbf{\bibinfo{volume}{380}},
  \bibinfo{pages}{151} (\bibinfo{year}{2001}), \eprint{astro-ph/0111092}.

\bibitem[{\citenamefont{{Wiringa} et~al.}(1988)\citenamefont{{Wiringa}, {Fiks},
  and {Fabrocini}}}]{1988PhRvC..38.1010W}
\bibinfo{author}{\bibfnamefont{R.~B.} \bibnamefont{{Wiringa}}},
  \bibinfo{author}{\bibfnamefont{V.}~\bibnamefont{{Fiks}}}, \bibnamefont{and}
  \bibinfo{author}{\bibfnamefont{A.}~\bibnamefont{{Fabrocini}}},
  \bibinfo{journal}{\prc} \textbf{\bibinfo{volume}{38}}, \bibinfo{pages}{1010}
  (\bibinfo{year}{1988}).

\bibitem[{\citenamefont{{M{\"u}ther} et~al.}(1987)\citenamefont{{M{\"u}ther},
  {Prakash}, and {Ainsworth}}}]{1987PhLB..199..469M}
\bibinfo{author}{\bibfnamefont{H.}~\bibnamefont{{M{\"u}ther}}},
  \bibinfo{author}{\bibfnamefont{M.}~\bibnamefont{{Prakash}}},
  \bibnamefont{and} \bibinfo{author}{\bibfnamefont{T.~L.}
  \bibnamefont{{Ainsworth}}}, \bibinfo{journal}{Physics Letters B}
  \textbf{\bibinfo{volume}{199}}, \bibinfo{pages}{469} (\bibinfo{year}{1987}).

\bibitem[{\citenamefont{{Lackey} et~al.}(2006)\citenamefont{{Lackey}, {Nayyar},
  and {Owen}}}]{2006PhRvD..73b4021L}
\bibinfo{author}{\bibfnamefont{B.~D.} \bibnamefont{{Lackey}}},
  \bibinfo{author}{\bibfnamefont{M.}~\bibnamefont{{Nayyar}}}, \bibnamefont{and}
  \bibinfo{author}{\bibfnamefont{B.~J.} \bibnamefont{{Owen}}},
  \bibinfo{journal}{\prd} \textbf{\bibinfo{volume}{73}}, \bibinfo{eid}{024021}
  (\bibinfo{year}{2006}), \eprint{astro-ph/0507312}.

\bibitem[{\citenamefont{Typel and Wolter}(1999)}]{Typel:1999yq}
\bibinfo{author}{\bibfnamefont{S.}~\bibnamefont{Typel}} \bibnamefont{and}
  \bibinfo{author}{\bibfnamefont{H.~H.} \bibnamefont{Wolter}},
  \bibinfo{journal}{Nucl. Phys.} \textbf{\bibinfo{volume}{A656}},
  \bibinfo{pages}{331} (\bibinfo{year}{1999}).

\bibitem[{\citenamefont{Coughlin et~al.}(2019)\citenamefont{Coughlin, Dietrich,
  Margalit, and Metzger}}]{Coughlin:2018fis}
\bibinfo{author}{\bibfnamefont{M.~W.} \bibnamefont{Coughlin}},
  \bibinfo{author}{\bibfnamefont{T.}~\bibnamefont{Dietrich}},
  \bibinfo{author}{\bibfnamefont{B.}~\bibnamefont{Margalit}}, \bibnamefont{and}
  \bibinfo{author}{\bibfnamefont{B.~D.} \bibnamefont{Metzger}},
  \bibinfo{journal}{Mon. Not. Roy. Astron. Soc.}
  \textbf{\bibinfo{volume}{489}}, \bibinfo{pages}{L91} (\bibinfo{year}{2019}),
  \eprint{1812.04803}.

\bibitem[{\citenamefont{Dietrich et~al.}(2020)\citenamefont{Dietrich, Coughlin,
  Pang, Bulla, Heinzel, Issa, Tews, and Antier}}]{Dietrich:2020lps}
\bibinfo{author}{\bibfnamefont{T.}~\bibnamefont{Dietrich}},
  \bibinfo{author}{\bibfnamefont{M.~W.} \bibnamefont{Coughlin}},
  \bibinfo{author}{\bibfnamefont{P.~T.} \bibnamefont{Pang}},
  \bibinfo{author}{\bibfnamefont{M.}~\bibnamefont{Bulla}},
  \bibinfo{author}{\bibfnamefont{J.}~\bibnamefont{Heinzel}},
  \bibinfo{author}{\bibfnamefont{L.}~\bibnamefont{Issa}},
  \bibinfo{author}{\bibfnamefont{I.}~\bibnamefont{Tews}}, \bibnamefont{and}
  \bibinfo{author}{\bibfnamefont{S.}~\bibnamefont{Antier}}
  (\bibinfo{year}{2020}), \eprint{2002.11355}.

\bibitem[{\citenamefont{Abrahamyan et~al.}(2012)}]{Abrahamyan:2012gp}
\bibinfo{author}{\bibfnamefont{S.}~\bibnamefont{Abrahamyan}}
  \bibnamefont{et~al.}, \bibinfo{journal}{Phys. Rev. Lett.}
  \textbf{\bibinfo{volume}{108}}, \bibinfo{pages}{112502}
  (\bibinfo{year}{2012}), \eprint{1201.2568}.

\bibitem[{\citenamefont{Essick et~al.}(2021)\citenamefont{Essick, Landry,
  Schwenk, and Tews}}]{Essick_skin}
\bibinfo{author}{\bibfnamefont{R.}~\bibnamefont{Essick}},
  \bibinfo{author}{\bibfnamefont{P.}~\bibnamefont{Landry}},
  \bibinfo{author}{\bibfnamefont{A.}~\bibnamefont{Schwenk}}, \bibnamefont{and}
  \bibinfo{author}{\bibfnamefont{I.}~\bibnamefont{Tews}},
  \emph{\bibinfo{title}{Astrophysical constraints on the symmetry energy and
  the neutron skin of $^{208}$pb with minimal modeling assumptions}},
  \bibinfo{howpublished}{\url{https://dcc.ligo.org/LIGO-P2100054}}
  (\bibinfo{year}{2021}).

\end{thebibliography}

\end{document}